\documentclass{aa}
\usepackage[utf8]{inputenc}

\usepackage{savesym}
\usepackage{amsmath}
\savesymbol{iint}
\usepackage[varg]{txfonts}
\restoresymbol{TXF}{iint}
\usepackage{graphicx}
\usepackage{units}
\usepackage{threeparttable} % für footnotes in tables
\usepackage{siunitx}
\usepackage{natbib}
 \usepackage{longtable}
 \usepackage{pdflscape} % for 'landscape' environment

  \renewenvironment{thebibliography}[1]{%
    \begin{oldthebibliography}{#1}%
      \setlength{\parskip}{0ex}%
      \setlength{\itemsep}{0ex}%
  }%
  {%
    \end{oldthebibliography}%
  }

\newcommand{\tx}[1]{\mathrm{#1}} 
\newcommand{\Steph}[1]{#1}

\usepackage{color}

\begin{document}

\title{Precise radial velocities of giant stars}

\subtitle{X. Bayesian stellar parameters and evolutionary stages for 372 giant stars from the Lick planet search\thanks{\Steph{Tables A.1 and A.2 are only available in electronic form
at the CDS via anonymous ftp to cdsarc.u-strasbg.fr (130.79.128.5)
or via http://cdsweb.u-strasbg.fr/cgi-bin/qcat?J/A+A/}
}}

\author{Stephan Stock 
\and Sabine Reffert \and Andreas Quirrenbach}

\institute{Landessternwarte, Zentrum für Astronomie der Universität Heidelberg, Königstuhl 12, 69117, Heidelberg, Germany}

\date{Received <day month year> /
Accepted <day month year>}

\abstract
{The determination of accurate stellar parameters of giant stars is essential for our understanding of such stars in general and as exoplanet host stars in particular. Precise stellar masses are vital for determining the lower mass limit of potential substellar companions with the radial velocity method, but also for dynamical modeling of multiplanetary systems and the analysis of planetary evolution.}{Our goal is to determine stellar parameters, including mass, radius, age, surface gravity, effective temperature and luminosity, for the sample of giants observed by the Lick planet search. Furthermore, we want to derive the probability of these stars being on the horizontal branch \Steph{(HB)} or red giant branch (RGB), respectively.}{We compare spectroscopic, photometric and astrometric observables to grids of stellar evolutionary models using Bayesian inference.}{We provide tables of stellar parameters, probabilities for the current post-main sequence evolutionary stage, and probability density functions for 372 giants from the Lick planet search. We find that $81\%$ of the stars in our sample are more probably on the \Steph{HB}. In particular, this is the case for 15 of the 16 planet host stars in the sample. We tested the reliability of our methodology by comparing our stellar parameters to literature values and find very good agreement. Furthermore, we created a small test sample of 26 giants with available asteroseismic masses and evolutionary stages and compared these to our estimates. The mean difference of the stellar masses for the 24 stars with the same evolutionary stages by both methods is only $\langle\Delta M\rangle=\unit[0.01\pm0.20]{M_\odot}$.}{We do not find any evidence for large systematic differences between our results and estimates of stellar parameters based on other methods. In particular we find no significant systematic offset between stellar masses provided by asteroseismology to our Bayesian estimates based on evolutionary models.}

\maketitle

\titlerunning{Precise radial velocities of giant stars}
\authorrunning{Stock \& Reffert}

% subdivision in 
%\section{Title}
%\subsection{Title}
%\subsubsection{Title}
%\paragraph{Title}

\section{Introduction}
Since the first discovery of an exoplanet around a main sequence star in 1995 by \cite{Mayor1995}, the number of detected extrasolar planets increased continuously. Most detected extrasolar planets are orbiting main sequence stars; however, a small number of planets have also been found around more evolved stars. The first planet around a giant star was discovered around $\iota$ Draconis by \cite{Frink2002}. Since then the number of detected substellar companions around giant stars has grown significantly to more than 100. The advantage of measuring radial velocities of giant stars is that they are cooler and rotate slower than their main sequence predecessors. This allows precise measurement of radial velocities from their spectra which would be very difficult for main sequence stars with stellar masses above $\unit[1.5]{M_\odot}$, as they have fewer absorption lines. As a result, by determining radial velocities of giant stars one can probe the planet population of more massive stars.

However, radial velocities of giant stars are subject to intrinsic radial velocity variability and jitter. Late-type K giants in particular can show large RV variations \citep{Setiawan2004, Hekker2006_2, Hekker2008}. These variations are caused by p-mode pulsations and occur on short timescales with periods ranging from several hours to days. Furthermore, \cite{Hatzes2004} postulates long-term variations of several hundred days that might be caused by non-radial pulsations. The latter could mimic radial velocity signals of extrasolar planets; see, for example,\Steph{ \cite{Hrudkova2017}, \cite{Hatzes2018} and \cite{Reichert}}. The knowledge of the host star's stellar parameters and current evolutionary stage is important for understanding such variations as well as for producing a clean planet sample around giant stars for a statistical analysis of planet occurrence rates around stars more massive than the Sun.
Furthermore, stellar parameters are essential for numerous other applications regarding planet-hosting stars. Some examples are the determination of the planet's minimum mass using radial-velocity data, dynamical modeling of multi-planetary systems, and analysis of planetary evolution around giant stars. For the latter, the current evolutionary stage is especially important as it provides the position of the planet around the host star at the time when the star was on the main sequence \citep{Kunitomo2011,Villaver2014,Currie2009}. This can help to understand planet-formation mechanisms \citep{Currie2009}. 

There is also an essential reason to know the current post-main sequence evolutionary stage of the giant star, when using the Hertzsprung-Russel diagram (HRD) or analogously the Color-Magnitude diagram (CMD) to determine stellar parameters from photometry and spectroscopy. The post-main sequence evolutionary tracks of evolved stars are degenerate in those diagrams, which allows for multiple solutions of stellar parameters of the giant star in these regions, depending on whether the star is on the red-giant branch (RGB), burning hydrogen in a shell, or on the horizontal branch (HB), burning helium in its core. 

Recently, some authors questioned the reliablity of stellar masses for giants stars determined via evolutionary models; see, for example, \cite{Lloyd2011}, \cite{Lloyd2013}, \cite{SchlaufmanWin2013} and \cite{Sousa2015}. These authors stated that stellar masses of giant stars determined from evolutionary models can be overestimated by a factor of two or even more. \cite{Takeda2015} argued that this bias towards higher masses might be caused by the fact that the RGB models cover
the whole area in the HRD which is occupied by
giant stars, whereas the HB models cover only part of that area. With the interpolation method, it is therefore always possible to find a solution on the RGB, but not necessarily on the HB, and the RGB solution for a star usually corresponds to a higher mass than the HB solution. \cite{Ghezzi2015} did not find any evidence for systematically higher masses from spectroscopic observations compared to either binary or asteroseismic reference masses. More recently, \cite{North2017} did not find any significant difference between asteroseismic masses and masses determined from spectroscopic observations in the range of $\unit[1]{M_\odot}$ to $\unit[1.7]{M_\odot}$, while \cite{Stello2017} find that spectroscopically determined masses for stars above $\unit[1.6]{M_\odot}$ can be overestimated by $15\%$-$20\%$. This overestimation could be caused by the simplifications that are used to calculate stellar evolutionary models. The assumption of local thermodynamic equilibrium (LTE) as well as the one-dimensional (1D) simplification might lead to significant uncertainties in stellar evolutionary models, especially regarding evolved stars \citep{Lind2012}.
A systematic overestimation of stellar masses across the whole mass range occupied by the Lick giant star sample would have consequences for the location of the peak in the planet occurrence rate as a function of stellar mass as obtained in \cite{Reffert2015}.

The purpose of this paper is to use a methodology based on Bayesian inference instead of interpolation to determine the stellar parameters of the giant stars in our sample as accurately as possible given the limitations of the observations and stellar evolutionary models. Furthermore, the Bayesian inference method can provide a probability estimate of the current evolutionary stage based on the likelihood of the models given the observed parameters and some physical prior information. It is therefore possible to more accurately asses the degeneracy of the two possible solutions of stellar parameters, corresponding to RGB and HB models, for many of our giant stars. \Steph{We also compare our results to those for other samples to evaluate the reliability of our methodology and stellar parameters.}

The paper is organized as follows. In Sect.~\ref{Sample}, our sample of stars for which we derived stellar parameters is described. Section~\ref{StellarModels} presents the stellar evolutionary models and their preparation for our application of the method, while Sect.~\ref{Methodology} explains the methodology for the stellar-parameter determination in detail. \Steph{In Sect.~\ref{Results}, we present the results of our sample and compare them to available literature values.} In Sect.~\ref{Discussion} we discuss the reliability of our estimations of stellar parameters, in particular stellar masses and evolutionary stages, by comparing them to asteroseismic test samples. Section~\ref{Summary} provides a short summary of the paper. 

\section{Sample}
\label{Sample}
\subsection{Stellar parameters}

\begin{table}
\caption{Adopted [Fe/H] values and their reference for the eight stars of our sample that have no metallicities in \cite{Hekker2007}.}
\label{tab: fehstars}
\centering
\begin{tabular}{r S[table-format=3.3(3),detect-weight,mode=text] l}
\hline\hline
HIP & {[Fe/H]} & Reference \\
\hline
476 &-0.020\pm0.050& \cite{feuillet2016}\\
4463 &0.050\pm0.060& \cite{Wu2011}\\
14915 &0.073\pm0.003& \cite{Ness2016}\\
46457 &0.070\pm0.080& \cite{Hansen1971}\\
67057 &0.110\pm0.110& \cite{Giridhar1997}\\
67787 &0.080\pm0.090& \cite{Franchini2004}\\
89918 &-0.170\pm0.090& \cite{Wu2011}\\
117503&-0.110\pm0.110& \cite{Wu2011}\\

\hline
\end{tabular}
\end{table}

Our sample of stars consists of 373 very bright ($V\le \unit[6]{mag}$) G- and K-giant stars with parallax measurements by Hipparcos \citep{Perryman1997a}. Their detailed selection criteria are outlined in \cite{Frink2001} and more recently in \cite{Reffert2015}. Selection criteria were a visual magnitude brighter than $\unit[6]{mag}$ and a small photometric variability. The sample started with 86 K giants in June 1999 and was extended with 93 K giants in 2000. In 2004, 194 stars that mostly belong to the G-giant regime were added to the sample. These stars have bluer colors (we used $0.8\leq B-V \leq 1.2$ as a selection criterion) and more importantly higher masses. The masses where roughly estimated at this time using evolutionary models by \cite{Girardi2000} with solar metallicity, as no individual metallicities were available. Only stars that were above the evolutionary track of $\unit[2.5]{M_\odot}$ were added to the sample. 

The radial velocities of these stars were monitored for more than a decade (1999-2011) with the $\unit[60]{cm}$ CAT-Telescope at Lick Observatory using the Hamilton Echelle Spectrograph ($R\sim60,000$). This resulted in several published planet detections; for example, \Steph{\cite{Frink2002}, \cite{Reffert2006}, \cite{Quirrenbach2011}, \cite{Mitchell2013}, \cite{Trifonov2014}, \cite{Ortiz2016}, \cite{Tala2018} and \cite{Luque2018}.}

For the derivation of stellar parameters for our sample, we used the $V$ and $B-V$ photometry provided by the Hipparcos catalog \citep{Hipparcos1997} together with spectroscopically determined metallicities by \cite{Hekker2007}. The Johnson $V$ and $B-V$ photometry of the Hipparcos catalog were determined from several sources. The Johnson $V$ magnitudes are either directly observed values from ground or transformations from the Hipparcos $H_p$ or Tycho $V_T$ magnitudes, whichever yielded a better accuracy \citep{Hipparcos1997}. The Johnson $B-V$ color index was derived using ground-based observations or the Tycho $B_T$ and $V_T$ magnitudes with the correct transformations corresponding to the luminosity class. \Steph{If the luminosity class of the star in the Hipparcos catalog was unknown, the transformations for luminosity class III giants were used \citep{Hipparcos1997}. This is favorable for our sample of giant stars, as we do not need to apply further corrections for these stars.} For some dozen stars in the Hipparcos catalog, no error of the $B-V$ color index was cataloged. For these stars we used the error given by the Hipparcos input catalog \citep{Turon1993}.

In addition to the $B-V$ color index, the Hipparcos catalog provides the $V-I$ color indices derived by various methods, which are described in detail in \cite{Hipparcos1997}. The reason for including the $V-I$ color index is that it helps to disentangle the degeneracy in the $B-V$ color index between M giants and late-type K giants, due to differential absorption by titanium oxide (TiO) in the stellar atmospheres of M giants, which affects the $V$ magnitude more than the $B$ magnitude.

\begin{figure}
\resizebox{\hsize}{!}{\includegraphics{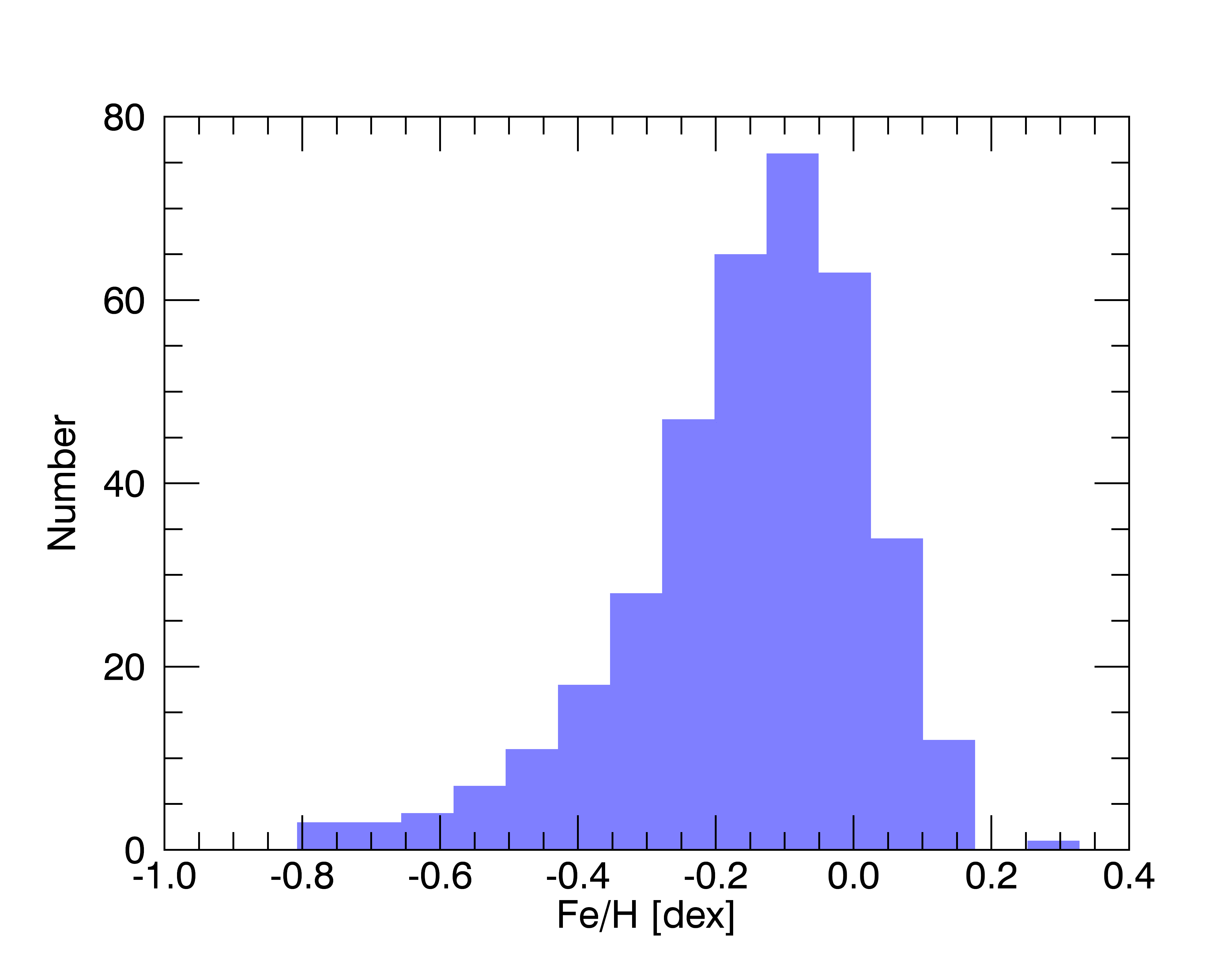}}
\caption{Histogram which shows the distribution of [Fe/H] values for our sample. }
\label{metfig}
\end{figure}

Since our sample of giant stars is part of a radial velocity survey to detect exoplanets, high resolution spectra are available for all stars.
Spectroscopic metallicities, effective temperatures, and surface gravities were determined by \cite{Hekker2007}. For the metallicities,
\cite{Hekker2007} estimated an error of $\sigma_{[Fe/H]}=\unit[0.1]{dex}$ by comparing some of their results to previous published literature values.
For eight stars of our sample, no metallicities could be determined from the spectra by \cite{Hekker2007}. For these stars we used the literature values shown in Table~\ref{tab: fehstars}. The mean metallicity of our sample is $\text{[Fe/H]}_{\text{mean}}=-0.116\; \text{dex}$. Figure~\ref{metfig} shows the metallicity distribution of our sample. In addition to the metallicites,  \cite{Hekker2007} determined surface gravities and effective temperatures with uncertainties of $\sigma_{log(g)}=\unit[0.22]{dex}$ and $\sigma_{\text{Teff}}=\unit[84]{K}$ for each star. While these two parameters could also be used to determine stellar parameters, for example, mass and radius,  from evolutionary models, we decided not to follow this approach. The reason is that spectroscopic data are often affected by relatively large unknown systematics, depending on the adopted models. However, we used the spectroscopic effective temperatures and surface gravity to verify the validity of our derived stellar parameters and we used the spectroscopic effective temperature in addition to the $V-I$ color index to break the degeneracy in the $B-V$ color index between M giants and late K giants.

\subsection{Parallax, distance, absolute magnitude and astrometry-based luminosity}

\begin{figure}
\resizebox{\hsize}{!}{\includegraphics{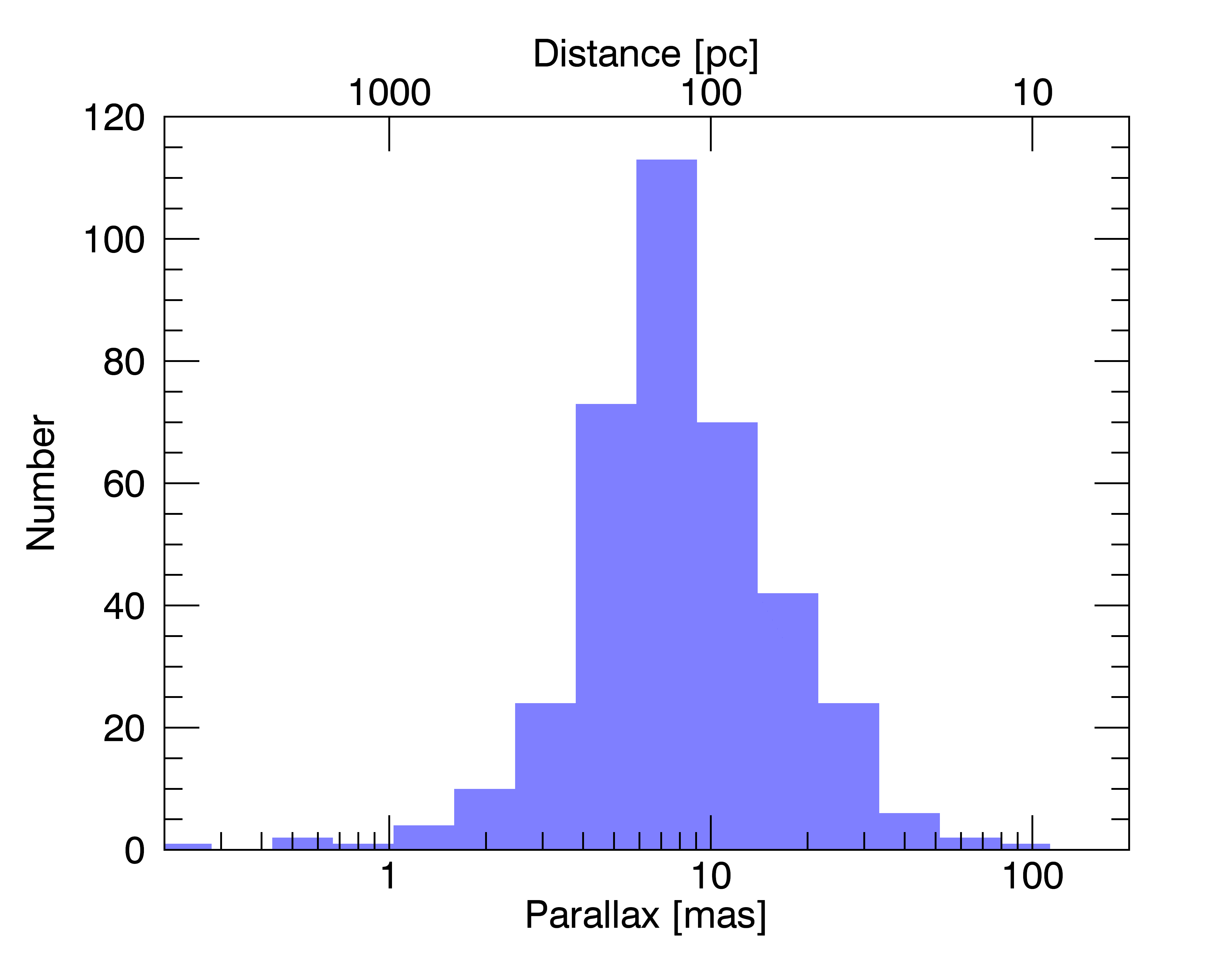}}
\caption{Histogram showing the distribution of measured parallaxes for our sample of stars. The x-axis is plotted on a logarithmic scale. For convenience, the upper x-axis shows the corresponding distance. }
\label{parfig}
\end{figure}

Comparing observed stars to evolutionary models requires information about the distances of the stars. 
Trigonometric parallax measurements by Hipparcos are available for all of the 373 giant stars studied here. We used the new reduction by \cite{vanLeeuwen2007} which has smaller statistical errors than the original Hipparcos catalog \cite{Hipparcos1997}.
The application of the distance modulus to determine absolute magnitudes from trigonometric parallaxes can lead to biases and systematic effects, due to the reciprocal and logarithmic transformation; see, for example, \citet{Lutz1973}, \cite{Arenou1999}, \cite{bailer2015} and \cite{bailer2016}.

Instead of the absolute magnitude we used the astrometry-based luminosity (ABL) introduced by \citet{Arenou1999} to compare observations with models. The ABL $a_\lambda$ at a certain wavelength $\lambda$ or in a certain photometric band can be derived by rewriting the distance modulus. The ABL is given by

\begin{equation}
\label{astrometriclum}
a_\lambda\equiv\varpi[\arcsec]\cdot10^{0.2m_\lambda[\text{mag}]-0.2A_\lambda[\text{mag}]+1}=10^{0.2M_\lambda[\text{mag}]}
,\end{equation}
where $\varpi$ is the trigonometric parallax (in arcseconds), $m_\lambda$ the apparent magnitude, $M_\lambda$ the absolute magnitude and $A_\lambda$ the extinction \citep{Arenou1999}. The ABL can be regarded as the reciprocal of the square root of the flux. In the case of our sample, the influence of the photometric error on the error of the ABL is several magnitudes smaller than the error due to the parallax, which is the reason why we neglected the photometric error for the determination of the error of the ABL. The advantage of the ABL is the linearity with the parallax, which implies that the ABL has Gaussian distributed symmetrical errors based on the errors of the parallax measurement. Furthermore, no biases, for example, Lutz-Kelker bias \citep{Lutz1973}, are introduced, and one could in principle also use negative parallaxes; fortunately in our sample of stars, no negative parallaxes are present. 

Figure~\ref{parfig} shows the distribution of trigonometric parallaxes for our sample of stars from \cite{vanLeeuwen2007}. The median parallax of our sample is $\unit[9.47]{mas}$. Due to the small distance of most of the stars in our sample, we neglect interstellar reddening when determining the ABL of our stars from Eq.~\ref{astrometriclum}.

\begin{figure*}[ht!]
\centering
\includegraphics[width=18.5cm]{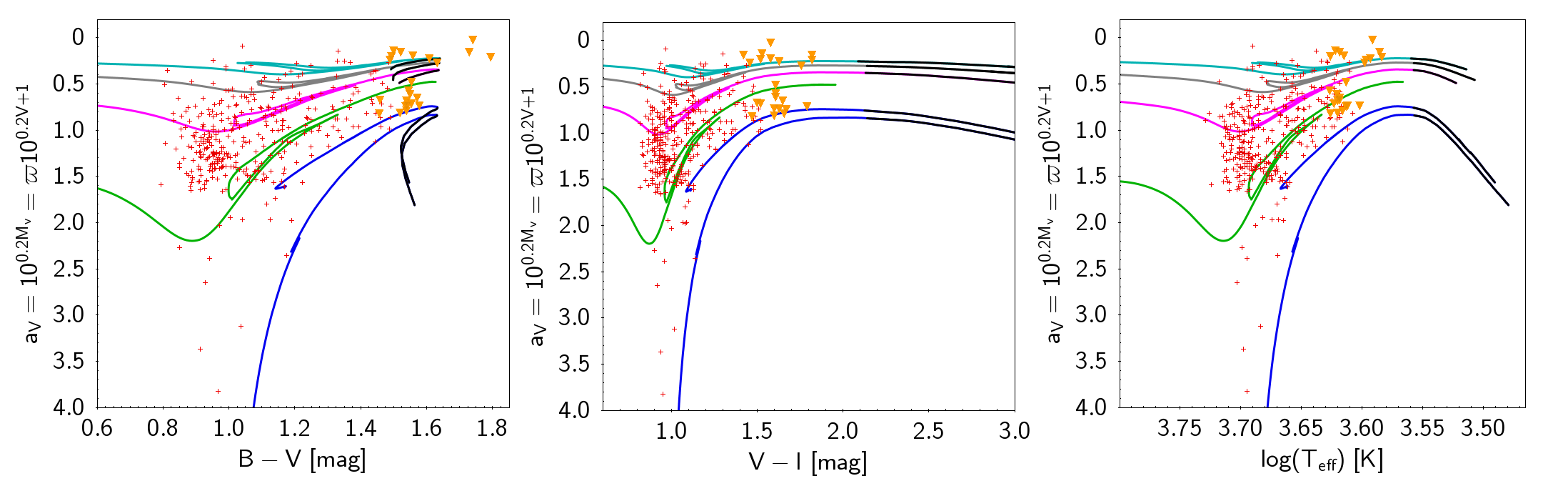}
\caption{Astrometric HRDs with different x-axes. From left to right: $B-V$, $V-I$ and $\log(T_{\text{eff}})$. The colored lines represent HB and RGB evolutionary tracks with metallicity Z=0.0180 and masses of $\unit[1]{M_\odot}$ (blue), $\unit[2]{M_\odot}$ (green), $\unit[3]{M_\odot}$ (violet), $\unit[4]{M_\odot}$ (gray) and $\unit[5]{M_\odot}$ (turquois). Bold black parts of the evolutionary track show the region that is affected by the degeneracy in $B-V$ due to the differential absorption of TiO bands in M giants, while orange triangles are stars that are in their 1-$\sigma$ range affected by this degeneracy. \Steph{The astrometric HRDs with $V-I$ or effective temperature plotted on the x-axis show that no giant star of our sample can be attributed to the M-giant region of the evolutionary tracks.}}
\label{AHR}
\end{figure*}

Figure~\ref{AHR} shows our sample of giant stars as red crosses in different so-called astrometric HRDs \citep{Arenou1999} with overplotted evolutionary tracks corresponding to a metallicity of Z=0.0180, which is very close to our derived mean metallicity of the sample. The degenerate part between K and M giants is marked in black on the evolutionary tracks and starts around $B-V\approx\unit[1.5]{mag}$ for each stellar model. Stars with $(B-V)+\sigma_{B-V}>\unit[1.5]{mag}$ are marked as orange triangles. Using the $V-I$ color index and the effective temperature to plot the astrometric HRD shows that these stars are not located in the M giant regime of the evolutionary tracks. This is especially apparent when using the $V-I$ color and its corresponding error estimate, as no star of our sample is in the 3-$\sigma$ range of the M giant regime (marked black) of the evolutionary tracks. This is why, later on,  we used the $V-I$ color index to exclude these models, thereby preventing artificial multiple solutions of stellar parameters for stars positioned in this region of the HRD that are purely based on the choice of photometry.

\section{Stellar evolutionary models}
\label{StellarModels}
\subsection{Models}
We determine stellar parameters using stellar evolutionary models provided by \citet{Bressan2012} based on the Padova and Trieste Stellar
Evolution Code (PARSEC), which include convective
overshooting. We chose the Padova models because of their widespread
use in the literature and the fact that they are available for a large range of stellar masses and metallicities. The mass grid ranges from $\unit[0.09]{M_\odot}$ to $\unit[12]{M_\odot}$ while the metallicity ranges from $Z=0.0001$ to $Z=0.06$. The helium content $Y$ and hydrogen content $X$ are regarded as known functions of $Z$ using the relation $Y=0.2485+1.78Z$. The assumed solar metal content for the models is $Z_\odot=0.0152$ \citep{Bressan2012}.
The models cover the range from the start of the pre-main sequence phase until the helium flash (HEF) for low-mass stars, to a few thermal pulses for intermediate-mass stars and until carbon ignition for massive stars \citep{Bressan2012}. Each evolutionary track includes values for the mass, radius, surface gravity, age, luminosity, effective temperature and the current evolutionary
phase. The evolutionary phase value is an increasing integer number at certain selected critical points
along the track, while fractional phase values are proportional to the fractional time duration between the previous critical point and the following one. Stars with masses $\unit[0.5]{M_\odot} < M  <  \text{M}_{\tx{HEF}}$, where M$_{\text{HEF}}$ depends slightly on the metallicity, experience a helium flash. For these low-mass stars, the evolutionary tracks are divided into the RGB and HB evolutionary phases. The evolution of the HB is computed
separately from a suitable zero-age horizontal branch (ZAHB) model which has the same core mass
and chemical composition as the last RGB model. The envelope mass of the HB models is regarded as a free parameter \citep{Bressan2012}, which is important to consider when adding mass loss along the RGB evolution.

\subsection{Interpolation to a finer grid of models}
Our derivation of stellar parameters requires a dense grid of stellar evolutionary models for different masses and metallicities. We interpolated a finer grid of stellar models using points of equivalent evolutionary phases of neighboring models. This is achieved by using the phase value provided by the PARSEC models. The evolutionary tracks are re-sampled so that they consist of exactly the same number of discrete models with the same phase values.
The advantage of this approach compared to other linear interpolations is that the shape and non-linearity of the stellar evolutionary tracks is conserved, especially for regions in the HRD where the evolutionary tracks undergo loops. 

Regarding the grid of stellar masses of the models, we interpolated to a grid of $\Delta M=\unit[0.025]{M_\odot}$ in the range between $\unit[0.5]{M_\odot}$ to $\unit[1.6]{M_\odot}$ and above $\unit[1.9]{M_\odot}$, while we interpolated to a mass grid of $\Delta M=\unit[0.05]{M_\odot}$ between $\unit[1.6]{M_\odot}$ and $\unit[1.9]{M_\odot}$ and $\Delta M=\unit[0.2]{M_\odot}$ between $\unit[10]{M_\odot}$ to $\unit[12]{M_\odot}$. We did not use the extension of the PARSEC library of stellar models with masses above $\unit[12]{M_\odot}$ provided by \cite{Tang2014} for several reasons: These models include a large amount of mass loss and have a discontinuity in their phase value compared to lower mass models. Both of these facts would necessitate a different preparation of the stellar models and their sampling if they were included. However, we do not expect many (if any) stars to have masses as high as $\unit[12]{M_\odot}$ in our sample. 

The coarser grid between $\unit[1.6]{M_\odot}$ and $\unit[1.9]{M_\odot}$ is due to the fact that it separates two physically different models of stars: low-mass stars that experience a helium flash and stars of higher masses that do not undergo a helium flash. The evolutionary tracks for these two types of stars have a slightly different shape and position in the HRD. An interpolation between these models can lead to non-physical and unreliable evolutionary tracks. For the determination of stellar parameters with our methodology we need a grid of stellar evolutionary tracks that is sampled equally in mass over the whole covered range of metallicities of the models; otherwise this will lead to artificial gaps in the determined probability density function (PDF) of the stellar mass. As a result, the precomputed coarser grid of stellar masses has to be kept in this mass range.

Regarding models of different metallicities, we interpolated to a grid that is 40 times finer than the precomputed grid provided by \cite{Bressan2012}. This was necessary for determining smooth probability density functions (PDFs) of the stellar parameters, especially for the stellar masses and ages, given the small observational errors of our sample.
Regarding models separated by $M_{\text{HEF}}$, the same difficulty as for the interpolation to a finer grid of stellar masses was encountered.
However, it is enhanced by the fact that M$_{\text{HEF}}$, and therefore the boundary between physically different models, varies over the range of metallicities. Therefore, it is not always possible to interpolate to a grid of at least $\Delta M=\unit[0.05]{M_\odot}$ between different metallicities. As a result, we interpolated all models for which this was possible in this mass range. We then filled the small number of gaps at certain masses and metallicities by extrapolating the stellar evolutionary tracks. This was achieved by using the two evolutionary tracks which have the closest stellar masses and the same metallicity as the evolutionary track that was extrapolated. With this approach no extrapolations by more than $\unit[0.05]{M_\odot}$ and only in very rare cases by $\unit[0.1]{M_\odot}$ were necessary. From inspection of the extrapolated evolutionary tracks in the HRD/CMD the inaccuracies due to the extrapolations are of the order of a few per cent at most. Due to the small number of extrapolations, and given the observational errors, modeling uncertainties due to small extrapolations are regarded as negligible.

\subsection{Bolometric corrections}

For the comparison of stellar evolutionary models to observed quantities we used bolometric corrections for the $V$ band and color-temperature relations for $B-V$ and $V-I$ by \cite{LeeandWorthey}. They provide extensive tabulated empirical relations for a large parameter space, from which one can determine the color and bolometric corrections by using a trilinear interpolation in metallicity, effective temperature and surface gravity. The absolute bolometric magnitude of the Sun $M_{\text{bol},\odot}$ cannot be set arbitrarily but has to be consistent with the tabulated bolometric corrections \citep{torres2010}. \cite{LeeandWorthey} adopted a solar bolometric correction of $BC_{\text{V},\odot}=-0.09$ which, together with our adopted solar absolute magnitude of $M_\text{V}=\unit[4.81]{mag}$ \citep{Hayes1985, torres2010}, results in $M_{\text{bol},\odot}=4.72$\,mag for the Sun. It is then straightforward 
to determine the absolute magnitudes for a given photometric band for each evolutionary model. 

\subsection{Further preparations}
The observed metallicities of the stars are measured in [Fe/H], while the model metallicities are given as fractional percentages $Z$. For the transformation of $Z$ to [Fe/H] we applied
\begin{equation}
[\tx{Fe/H}]=\log\frac{X_\odot}{Z_\odot}\cdot\frac{Z_\star}{X_\star}.
\end{equation}
We use [Fe/H] to denote the whole metal content of the star. 

The evolutionary models by \cite{Bressan2012} do not explicitly give mass loss for stars below $\unit[13]{M_\odot}$. We calculated the mass loss along the RGB for each stellar evolutionary model according to Reimer's law \citep{reimers75} using the modest value of $\eta=0.2$ as suggested by \citet{Miglio2012}. We used the corresponding HB models with the same core and envelope mass to follow up the evolution of the RGB, where the envelope mass is regarded as a free parameter based on the mass loss along the RGB. As the HB models are computed from a suitable zero-age horizontal branch they have an assigned starting age of zero. For the starting age of these models we used the age at the end of the corresponding RGB model. 

Very low-mass stars have main sequence lifetimes larger than the current age of the
Universe. Therefore, the evolutionary models were truncated when the age along the track reached the limit of 13.8 billion years. This results in a corresponding minimum mass of the post-main sequence models of around $\unit[0.75-0.9]{M_\odot}$ depending on the metallicity and evolutionary stage, that is, RGB or HB. Stars of lower masses have not evolved off the main sequence yet.

For the determination of the current evolutionary stage with our methodology we divided all models into RGB and HB models.

\section{Methodology}
\label{Methodology}
The Bayesian methodology we applied is very similiar to the methodologies outlined in \cite{Jorgendsen2005} and \cite{Silva2006}. However, some modifications and improvements were added. One fundamental difference is that \cite{Jorgendsen2005} and \cite{Silva2006} used isochrones instead of evolutionary tracks to determine stellar parameters. While both are in general equivalent to each other, they still need a different treatment when using Bayesian inference to determine stellar parameters.  
The following section explains how the Bayesian methodology was applied to determine stellar parameters of giant stars from stellar evolutionary tracks. 

\subsection{Determination of the posterior probability}
Each evolutionary track consists of discrete points that have defined stellar parameters and positions in a three-dimensional cube of metallicity, effective temperature, and luminosity. One may use alternative parameterizations for this cube; we use color $B-V$ and ABL $a_V$ instead of effective temperature and luminosity. 

We connected the discrete points along each evolutionary track to small sections, which we label $k$. Each of these sections is assigned the mean stellar parameter between the start ($k-\frac{1}{2}$) and the end point ($k+\frac{1}{2}$) of this section. Additionally, we assigned to each section an evolutionary time $\Delta \tau_{i,j,k}=t_{i,j,k+\frac{1}{2}}-t_{i,j,k-\frac{1}{2}}$, depending on the age at the start and end point.

For a star with observables that have Gaussian distributed errors one can estimate the likelihood of the star belonging to the section $k$ of the evolutionary track with mass $i$ and metallicity $j$ by
\begin{align}
\label{likelyh}
L_{i,j,k}\propto & \exp\bigg(-\frac{(a_V-a_{V_{i,j,k}}')^2}{\sigma_{a_V}^2}-\frac{((B-V)-(B-V)_{i,j,k}')^2}{\sigma_{B-V}^2}\\&-\frac{([\text{Fe/H}]-[\text{Fe/H}]_{j}')^2}{\sigma_{[\text{Fe/H}]}^2}\bigg),
\end{align}
where $a_{V_{i,j,k}}'$ and $(B-V)_{i,j,k}'$ are the mean values corresponding to the section $i,j,k$ of the model and $[\text{Fe/H}]_{j}'$ is the metallicity of the model.

The posterior probability is proportional to the product of the likelihood and the prior probability, meaning that

\begin{equation}
P_{i,j,k}\propto L_{i,j,k}\cdot\Delta n_{i,j,k}\cdot w_j
,\end{equation}
where $\Delta n_{i,j,k}$ is the number density of predicted stars which populate the evolutionary track section $k$ of mass $i$ and metallicity $j$ and $w_j$ the weight of models with metallicity $j$. It is essential to include the weight of the models, as they are not sampled equally in [Fe/H]. This is given by
\begin{equation}
w_{j}=\frac{\frac{1}{2}|[\text{Fe/H}]_{j}-[\text{Fe/H}]_{j-1}|+\frac{1}{2}|[\text{Fe/H}]_{j+1}-[\text{Fe/H}]_{j}|}{|[\text{Fe/H}]_{\text{max}}-[\text{Fe/H}]{_\text{min}}|}, 
\end{equation}
with $[\text{Fe/H}]_{\text{min}}$ and $[\text{Fe/H}]{_\text{max}}$ being the minimum and maximum metallicity of the models.

Regarding the complete derivation of $\Delta n_{i,j,k}$ we refer to \cite{Shull1993}. 
In short, the number of stars that populate a certain section of the evolutionary track at a certain age $t$ given a star formation rate (SFR) $\Psi (t)$ can be written as

\begin{equation}
\label{popev}
n_{i,j,k}(t)\propto N_{i,j}^0\int_{t_{i,j,k-\frac{1}{2}}}^{t_{i,j,k+\frac{1}{2}}} \Psi (t-t')\tx{d}t', 
\end{equation}
where $N_{i,j}^0$ is the number density of stars of a certain mass and metallicity that are instantaneously born, that is, it is the \Steph{initial mass function (IMF)} for a given metallicity. With the simplified assumption that the SFR was constant in the Galaxy, and due to the fact that we do not have a simple stellar population but field stars with different ages, one can determine the expected number density of stars by integrating Eq.~\ref{popev}. This results in
\begin{equation}
n_{i,j,k}(t)\propto N_{i,j}^0\cdot (t_{i,j,k-\frac{1}{2}}-t_{i,j,k+\frac{1}{2}}) =N_{i,j}^0\cdot\Delta \tau_{i,j,k}.
\end{equation}
As a result, the total prior distribution for field stars under the assumption of constant SFR in the Galaxy is given by
\begin{equation}
\label{prior}
\Delta n_{i,j,k}\propto \frac{\tx{d}n_{i,j}}{\tx{d}m_{i,j}}\cdot \Delta \tau_{i,j,k}=m_i^{-\alpha}\cdot\Delta \tau_{i,j,k}
,\end{equation}
where $\frac{\tx{d}n_{i,j}}{\tx{d}m_{i,j}}$ is the IMF and $\Delta \tau_{i,j,k}$ the evolutionary time.
Without the simplification of a constant SFR in the past one would need to take the total evolutionary flux along the evolutionary track into account \citep{Renzini1986}. This would lead to an additional factor $\frac{\tx{d}m_{i,j,k}}{\tx{d}t_{i,j,k}}$ which can be regarded as the death rate of the stellar population \citep{Renzini1986} but is not needed in our case.

Most of our giant stars have expected masses above $\unit[1]{M_\odot}$, which is the reason why we simply used the IMF by \cite{Salpeter1955} with $\text{d}N\propto M^{-\alpha} \text{d}M$ and $\alpha=2.35$. Due to the fact that our stellar models are sampled equally in stellar mass it is not important to integrate the IMF for each stellar mass range of the models. The stellar age along each evolutionary track however is not sampled equally. As a result, the evolutionary time $\Delta \tau_{i,j,k}$ serves not only as a prior that increases the probability for stars that are in a slower phase of their evolution, it also removes the dependency on the sampling. Without taking into account the evolutionary time, the posterior probability distribution would be biased towards denser sampled regions of the evolutionary tracks. 

The calculation of the posterior probability for each section of the fine grid of evolutionary models is computationally demanding. Therefore, we used only models that are within the 5$\sigma$ range of the measured position of the star in the astrometric HRD. Some tests have shown that including models further away has almost no influence on the resulting stellar parameters, but at a cost of a much higher computation time. 

The main improvement of our methodology compared to the approaches by \cite{Jorgendsen2005} or \cite{Silva2006} is the fact that we differentiate between RGB and HB models which are degenerate in the HRD. Because we separate these models, we can calculate the probability density function for each stellar parameter for each of the two evolutionary stages separately. Furthermore, this allows us to determine the overall posterior probability of each of these two cases given by
\begin{equation}
P_{l}=\frac{\sum_{l}{P_{i,j,k,l}}}{\sum_{l=0}{P_{i,j,k,l}}+\sum_{l=1}{P_{i,j,k,l}}}
,\end{equation}
where $l=0$ stands for the RGB and $l=1$ for the HB models.

\subsection{Probability density functions}
Each section $k$ of the evolutionary tracks with mass $i$, metallicity $j$ and posterior probability $P_{i,j,k}$ corresponds to a certain stellar parameter $X_{i,j,k}$ of the track; for example, radius. To determine the probability density functions (PDFs), we create histograms for each stellar parameter. In our case, we determine the PDFs for mass, radius, age, effective temperature, surface gravity and luminosity. Since our stellar masses of the evolutionary models are equally sampled over all models, the binning of the mass histogram is determined by the mass grid of $M=\unit[0.025]{M_\odot}$. The other stellar parameters change along each track and are irregularly sampled over all models. As a result, the other stellar parameters vary much more and can have many different values. For the construction of histograms for these irregularly sampled stellar parameters, we used an automated binning technique introduced and discussed by \cite{Hogg2008}, which is based on a jackknife likelihood. The automated binning technique adjusts the bin size and number of bins depending on the number of models $X_{i,j,k}$ and their posterior probability $P_{i,j,k}$. This is essential, as non-optimal binning would either result in loss of physical information or induce  artificial gaps in the PDF. In some cases where the probability of the evolutionary stage is very low and not many models are within the 5$\sigma$ range corresponding to the measured position of the star, the automated binning is not optimal. This is one of the reasons why less probable solutions tend to have more multi-modal PDFs. Therefore, we only provide stellar parameters and PDFs for solutions of an evolutionary stage that has a probability larger than $1\%$.

The resulting histograms show the underlying probability distribution, binned and on a relatively coarse grid. We normalize each histogram in such a way that the maximum is one. In order to determine a good estimate of the mode and mean of the underlying probability density function, we numerically smooth the histogram with a spline interpolation. In most cases, this results in a very good approximation of the underlying histogram. While the main cause for a possible multi-modal PDF is excluded due to our separation of the RGB and HB models, the PDF can still have multiple modes and in many cases it cannot be described by an analytic function such as a Gaussian. The reasons lies in the non-linearity of the evolutionary tracks and their curvature. This is most relevant for the beginning of the HB models, which, for low-mass stars, start with a loop at the region of the red clump, or, for higher-mass stars, at the so-called blue loop.  

\subsection{Stellar parameters and their confidence intervals}
\begin{table*}[t]
\begin{center}
\caption{Flags used in the catalog of stellar parameters.}
\label{Tab: Flags}
\begin{tabular}{l l}
\hline\hline
Flag& Meaning\\
\hline
V&$V-I$ color index used to determine stellar parameters (instead of $B-V$)\\
D&Degeneracy for late-type K giants resolved by applying a cut of the models based on the spectroscopic $T_{\text{eff}}$\\
L&RGB mass loss below threshold, no mass loss provided\\
P&Probability of evolutionary stage less than 1$\%$, no parameters provided\\
$X$M\tablefootmark{a}& Multi-modal PDF\tablefootmark{b}\\
$X$C\tablefootmark{a}& Mean of PDF is outside the 1$\sigma$ confidence interval\tablefootmark{c}\\
\hline
\end{tabular}
\tablefoot{
\tablefoottext{a}{$X$ represent the stellar parameter and can be $M$ (mass), $R$ (radius), $A$ (age), $T$ (temperature), $G$ (surface gravity) and $L$ (luminosity).}
\tablefoottext{b}{For example, the flag MM means that the PDF of the stellar mass has multiple modes and should be checked before blindly adopting the provided value.}
\tablefoottext{c}{For example, the flag AC means that the determined mean age from the PDF is outside the 1$\sigma$ confidence level}
}
\end{center}
\end{table*}

From the smoothed PDFs, one can determine the mean of the distribution and the most probable value, which is the maximum of the distribution (mode). The mode has the advantage that in cases of very broad PDFs, which are truncated, or in the case of multi-modal PDFs, it is less biased towards the center of the range of stellar parameters \citep{Jorgendsen2005}. Furthermore, in cases where the parameter estimation with the Bayesian methodology is uncertain, due to large observational errors, for example, the PDF will be very broad and possibly truncated. In these cases, the mode tends to be located at the extreme value of the stellar parameters, which often indicates that the parameter to be determined is not well constrained by the observables. While this problem is not particularly important for our sample of close giants with small observational errors, it is a problem for more distant stars or those with large observational errors in general. Unlike the mean, the mode is not a moment of the PDF, which is why we provide both values.

As many PDFs cannot be approximated by an analytic function, it is not straightforward to determine the confidence intervals of our stellar parameters. Let $f(X)$ be the PDF of the stellar parameter $X$. We calculate the confidence interval by
\begin{equation}
\label{CI1}
\frac{\int_{X_1}^{X_2}f(X) \tx{d}X}{\int_{X_{\tx{min}}}^{X_{{\tx{max}}}}f(X) \tx{d}X}=1-\alpha
,\end{equation}
where $X_{\tx{min}}$ and $X_{\tx{max}}$ are the extreme values as covered by the stellar evolutionary tracks, $X_1$ and $X_2$ are the lower and upper boundary for the confidence interval and $1-\alpha$ is the 1$\sigma$ (0.6827) or  3$\sigma$ (0.9973) confidence level. However, we additionally add the constraint that
\begin{equation}
\label{CI2}
f(X_1)=f(X_2),
\end{equation}
which leads to confidence intervals that allow a better description of non-symmetric PDFs.
Our confidence interval is always defined to include the mode of the PDF. For non-symmetrical or multi-modal PDFs, this can lead to a situation where the determined mean of the PDF is outside the 1$\sigma$ confidence interval. Stellar parameters from PDFs where this is the case should be regarded with caution.

This work provides a catalog of the stellar parameters for 372 giant stars \Steph{which is only available electronically at the CDS online archive} \footnote{\Steph{via anonymous ftp to cdsarc.u-strasbg.fr (130.79.128.5)
or via http://cdsweb.u-strasbg.fr/cgi-bin/qcat?J/A+A/
}}. We include the corresponding evolutionary stages and their probabilities as well as their associated stellar parameters, given as the mean and mode of the PDFs \footnote{\Steph{Send requests for the individual PDFs to S. Stock, e-mail: sstock@lsw.uni-heidelberg.de}}. We also provide the lower and upper 1$\sigma$ and 3$\sigma$ confidence limits and include a number of flags that summarize some properties of the provided stellar parameters, indicating, for example,\ when a PDF has multiple modes. The flags and their meaning are tabulated in Table~\ref{Tab: Flags}. A blank space for a stellar parameter in the catalog indicates that the determination was not successful. 

\section{Results and comparisons}
\label{Results}

\subsection{Stellar masses and evolutionary stages}
\label{evolution}
\begin{figure*}[ht!]
\centering
\includegraphics[width=18.5cm]{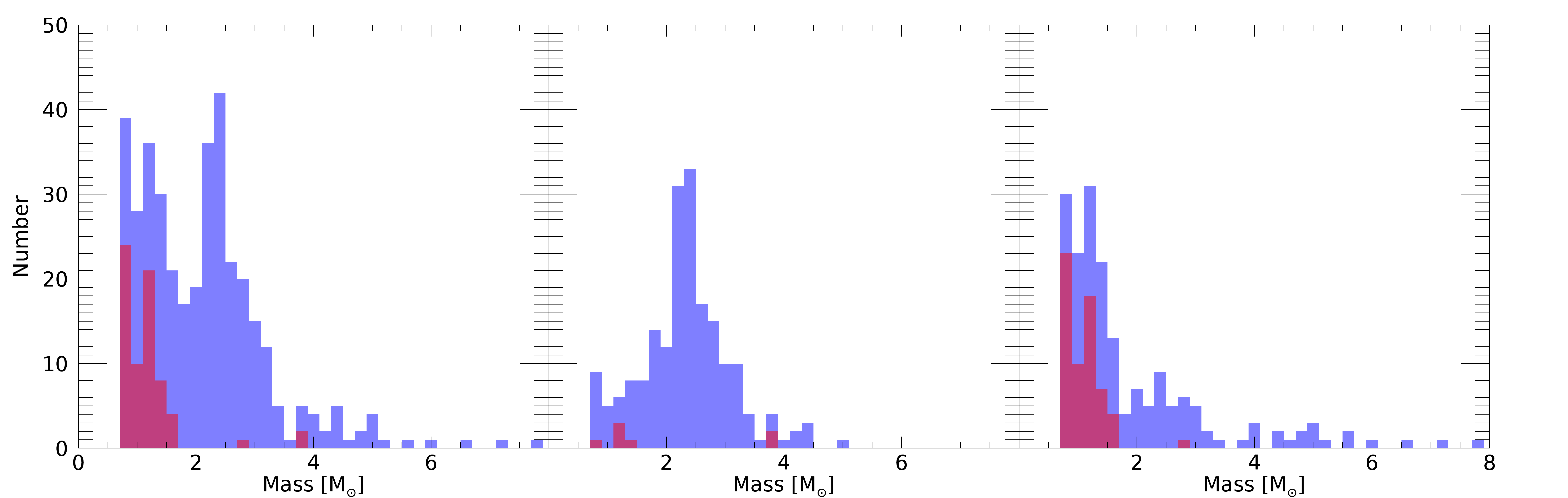}
\caption{\textit{Left}: Histogram representing the mass distribution of all stars in our sample, plotted in blue. Those stars that are estimated to be on the RGB by our method are plotted in red. \textit{Middle}: Mass distribution of the subsample that was added in 2004 to our sample of giants. One selection criterion of this sample was that the stars are positioned above the evolutionary track of $\unit[2.5]{M_\odot}$ for solar metallicity in the HRD. \textit{Right}: Subsample of stars which were not selected according to stellar mass.}
\label{Fig: Massdistro}
\end{figure*}

We were able to determine the stellar parameters for 372 of the 373 stars from the Lick planet search \citep{Frink2001, Reffert2015}. The one star for which we could not determine the stellar parameters is HIP 33152 (HD 50877). This star has a very small parallax ($\varpi=\unit[0.22\pm0.43]{mas}$) in the new reduction of the Hipparcos catalog \citep{vanLeeuwen2007}, resulting in a stellar mass that is probably higher than $\unit[12]{M_\odot}$ which is outside of our adopted grid of stellar evolutionary models. Due to the fact the parallax is not significantly different from zero and the star therefore probably suffers from significant extinction and reddening, we chose not to include the star in our catalog.

We find that 70 of the 372 giant stars ($\sim\!\!18.8\%$) are probably on the RGB while 302 stars ($\sim\!\!81.2\%$) are on the HB. The evolutionary stage is assigned based on its probability, which has to be higher than 50\% for one of the two possible stages. By stacking the mass PDFs of all the stars for the more probable evolutionary stage, we derive a mean mass of the sample of $\unit[2.7]{M_\odot}$. In Fig.~\ref{Fig: Massdistro} we show the mass distribution (blue) by adopting the mode as the mass value (without including the complete PDF) for all observed giants. The mean of the distribution of mode masses ($\unit[2.2]{M_\odot}$) is smaller than the mean mass derived by stacking the PDFs. The fraction of all giants that are more likely on the RGB is also shown in Fig.~\ref{Fig: Massdistro} by the red distribution. We outline several key features of the observed mass distribution. One can clearly see a peak for the HB stars between $\unit[2]{M_\odot}$ and $\unit[3]{M_\odot}$. There are several reasons for the appearance of such a peak in our sample of stars. One is that the evolutionary time of HB stars peaks between $\unit[1.7]{M_\odot}$ and $\unit[2]{M_\odot}$ due to the fact that these stars do not go through the helium flash 
\citep[see e.g.,\ Fig.~6 in][]{Girardi2013}. However, the RGB phase of such stars is very short, which explains the observed lack of RGB stars at masses higher than $\unit[1.7]{M_\odot}$. Regarding the evolutionary time scales and IMF one would expect to have a much larger number of RGB stars with masses around $\unit[1]{M_\odot}$, at least in a volume-limited sample. However, we emphasize that our sample is not volume-limited. More than half of the stars of our sample (194 stars that were added in 2004) were specifically chosen to have higher stellar masses. This subsample consists of stars that are located above the $\sim\!\!\unit[2.5]{M_\odot}$ solar metallicity evolutionary track by \cite{Girardi2000}. This approach was chosen in order to investigate the planet occurrence rate and planet masses around higher-mass host stars; for details, we refer to \cite{Reffert2015}. The mass distribution of the stars that are in this subsample of more massive stars is given in the middle panel of Fig.~\ref{Fig: Massdistro}. One can clearly see that the peak for the stellar masses around $\unit[2.5]{M_\odot}$ can be associated with this subsample. The right panel in Fig.~\ref{Fig: Massdistro} shows the remaining sample for which we did not apply stellar mass as a selection criterion. The distribution of this subsample is in closer agreement with a sample that is volume-limited, as the number of RGB stars increases towards lower masses. 

For statistical analyses and further comparison of our derived stellar parameters to literature values we use the following approach:
We provide the average difference of the stellar parameters determined by our method and a reference as well as their standard deviation. We usually adopt the mode of the PDF as our determined stellar parameter, without considering the confidence interval. 
We provide comparison plots of the two stellar parameters, where our Bayesian values are on the ordinate and the literature values on the abscissa. Additionally, we plot the difference (in case of stellar radii also the fractional difference) between the Bayesian and the reference value as a function of the reference value.

It can sometimes be difficult to choose between the\  mode and mean of the PDF when adopting a  stellar parameter. Normally this should be decided by looking at the PDF of each individual stellar parameter. We circumvent this ambiguity and also include the uncertainties of the stellar parameters in the comparison. To this end, we carry out a more in-depth analysis which takes the full PDFs of the stellar parameters into account. For our Bayesian values we use the smoothed PDFs as determined in this work, while for the reference values that have symmetric errors, we assume a normal distribution. For some reference values, namely those that have non-symmetric errors and no further information regarding the shape of their PDFs, we assume a split normal distribution \citep{John1982,Villani2006}.
We use a Monte-Carlo approach to sample the two-dimensional (2D) PDF of the stellar parameter comparison with $10,000$ synthesized points for each star. These points are drawn from the two PDFs by inverse transform sampling and with iteration. From these points, we estimate the PDF of the difference of both values for each star. 
This results in a total of $10,000$ single linear fits from which we determine the distribution of the slope and intercept by using a kernel density estimator with an Epanechnikov kernel \citep{Ep1969}.
The best fit is determined by the mode of the distributions for the slope and intercept. For the estimation of the errors, we calculate the confidence interval of the slope and intercept from their distributions by analogously applying Eqs.~\ref{CI1} and \ref{CI2}. We also visualize the significance of the best fit by a confidence band that includes $95\%$ of all fits. 

Unfortunately we do not have many stellar reference masses from model-independent methods like asteroseismology or binary mass measurements available. For this reason, we created a test sample to discuss the reliability of our mass estimates (see Sect.~\ref{Stellar_Mass_discussion}). However, we also compare our Bayesian stellar masses to previous mass estimates from \cite{Reffert2015} based on evolutionary tracks (see Sect.~\ref{mass_comp}).

\subsection{Stellar radii}
\label{stellar_radii}
\begin{table*}[tbp]
\caption{Best fit parameters for the comparison between Bayesian stellar parameters from evolutionary tracks to stellar parameters from the literature. }
\label{Tab: all}
\begin{center}

\begin{tabular}{l r c r@{}c@{}l r@{}c@{} lr@{}c@{} l}
\hline\hline
  \multicolumn{1}{c}{Sample} &
  \multicolumn{1}{c}{ $N_\text{Stars}$} &
  \multicolumn{1}{c}{Quantity} &
  \multicolumn{3}{c}{Value} &
  \multicolumn{3}{c}{$m$\tablefootmark{*}} &
  \multicolumn{3}{c}{ $c$\tablefootmark{*}} \\  
\hline\\ [-1.7ex]
\vspace{0.08cm}
Lick $\cap$ Charm2& 86&$\langle R_{\text{trk.}}/R_{\text{Ch.}}\rangle$ & 0.98&$\pm$&0.10& $0$&$.$&$\unit[000^{+0.001}_{-0.001}]{R_\odot^{-1}}$& $0$&$.$&$99^{+0.02}_{-0.02}$  \vspace{0.04cm} \\
Lick $\cap$ Charm2& 86&$\langle R_{\text{trk.}}-R_{\text{Ch.}}\rangle$&$-0.69$&$\pm$&$\unit[10.10]{R_\odot}$& $-0$&$.$&$04^{+0.15}_{-0.15}$& $0$&$.$&$\unit[89^{+3.94}_{-3.63}]{R_\odot}$  \vspace{0.04cm} \\
Lick $\cap$ \cite{Hekker2007}& 364&$\langle T_{\text{eff, trk.}}-T_{\text{eff, spec.}}\rangle$&$-41$&$\pm$&$\unit[106]{K}$ & $-0$&$.$&$02^{+0.02}_{-0.01}$&$43$&$.$&$\unit[94^{+52.47}_{-81.10}]{K}$  \vspace{0.04cm} \\
Lick $\cap$ \cite{Hekker2007}& 364&$\langle\log(g)_{\text{trk.}}-log(g)_{\text{spec.}}\rangle$&$-0.36$&$\pm$&$\unit[0.26]{dex}$ & $-0$&$.$&$\unit[17^{+0.02}_{-0.02}]{}$& $0$&$.$&$\unit[09^{+0.06}_{-0.05}]{dex}$ \vspace{0.04cm} \\
Lick $\cap$ \cite{Reffert2015}& 361&$\langle M_{\text{trk.}}-M_{\text{p.}} \rangle$ &$-0.12$&$\pm$&$\unit[0.47]{M_\odot}$&$-0$&$.$&$04^{+0.04}_{-0.03}$&$-0$&$.$&$\unit[05^{+0.07}_{-0.09}]{M_\odot}$ \vspace{0.04cm} \\
Lick $\cap$ \cite{Reffert2015} \tablefootmark{a}& 202&$\langle M_{\text{trk.}}-M_{\text{p.}} \rangle$ & $0.00$&$\pm$&$\unit[0.48]{M_\odot}$ &$0$&$.$&$02^{+0.05}_{-0.04}$&$-0$&$.$&$\unit[06^{+0.08}_{-0.11}]{M_\odot} $ \vspace{0.04cm}\\
Lick $\cap$ \cite{Reffert2015}\tablefootmark{b}& 159&$\langle M_{\text{trk.}}-M_{\text{p.}} \rangle$ & $-0.27$&$\pm$&$\unit[0.42]{M_\odot}$&$-0$&$.$&$24^{+0.06}_{-0.04}$ &$0$&$.$&$\unit[28^{+0.09}_{-0.15}]{M_\odot}$\vspace{0.04cm}\\
Test Sample\tablefootmark{c}& 24&$\langle M_{\text{trk.}}-M_{\text{Ast.}}\rangle$&$0.01$&$\pm$&$\unit[0.20]{M_\odot}$&$-0$&$.$&$21^{+0.14}_{-0.11}$ &$0$&$.$&$\unit[36^{+0.17}_{-0.21}]{M_\odot}$\vspace{0.04cm} \\
Test Sample\tablefootmark{d}& 24&$\langle M_{\text{trk.}}-M_{\text{Ast.}}\rangle$&$0.01$&$\pm$&$\unit[0.20]{M_\odot}$&$-0$&$.$&$03^{+0.16}_{-0.14}$ &$0$&$.$&$\unit[05^{+0.23}_{-0.23}]{M_\odot}$\vspace{0.04cm} \\

\hline\\ [-1.7ex]
\hline
\end{tabular}
\tablefoot{
\tablefoottext{*}{coefficient of linear model $y=mx+c$}
\tablefoottext{a}{stars that have been assigned the same evolutionary stage}
\tablefoottext{b}{stars that have a different assigned evolutionary stage}
\tablefoottext{c}{fit with $x$ and $y$ errors}
\tablefoottext{d}{fit with $y$ error only}
}
\end{center}
\end{table*}

Eighty-six stars of our sample of giant stars have entries in the CHARM2 catalog by \cite{Richichi2005} with limb darkening corrected angular diameters. For these we derived the stellar radii $R_{\text{Ch.}}$ with the help of the parallaxes from \cite{vanLeeuwen2007}. We propagated the non-Gaussian errors that arise from transforming trigonometric parallaxes and limb-darkened disk diameters to stellar radii by using a Monte-Carlo approach similar to the one explained above. This allows us to derive the PDFs of the stellar radii from CHARM2. Due to the fact that the stellar radii of our sample of stars span two orders of magnitude, we did not only calculate the statistics for the absolute differences of both values, but also for their fraction. We show the comparison of these values in Fig.~\ref{Fig: all} and the corresponding statistical parameters of the comparison in Table~\ref{Tab: all}. We find a mean ratio that is close to one, well within its standard deviation. The linear fit parameters are within their 1$\sigma$ range also consistent with a slope and offset of zero. Regarding the absolute difference we see that the radii derived from the mode of the PDF $R_{\text{trk.}}$ are on average $\langle R_{\text{trk.}}-R_{\text{Ch.}}\rangle=\unit[-0.69\pm10.10]{R_\odot}$ smaller than the estimated radii from CHARM2. However, the standard deviation is one order of magnitude larger. The resulting parameters of the linear fit that takes the complete PDFs into account do not show a significant slope or offset.
We therefore do not find significant systematic differences and conclude that our stellar radii are good estimates of the true stellar radii. 

\subsection{Effective temperatures}

\begin{figure*}
\centering
\includegraphics[width=15.5cm]{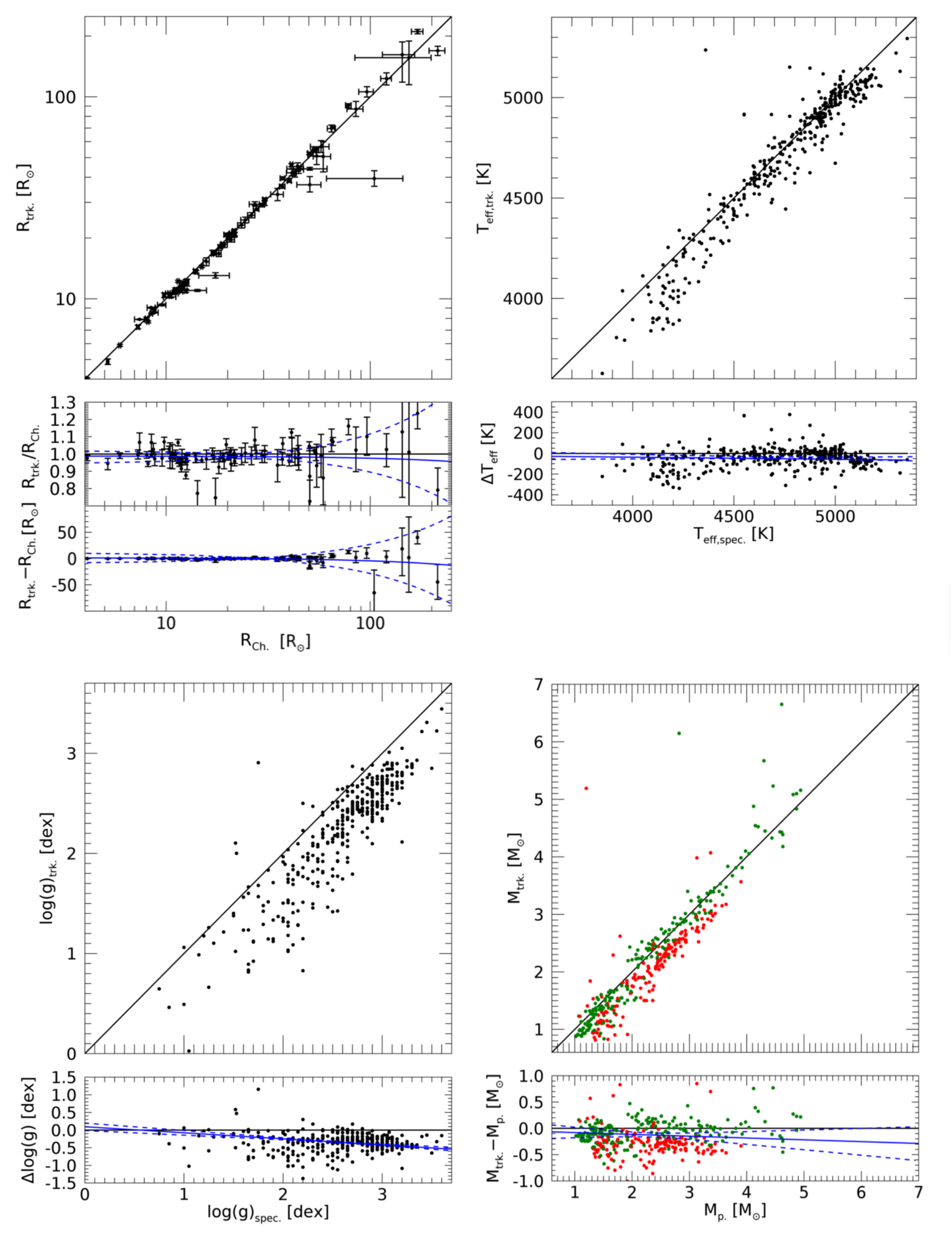}
\caption{Stellar parameters for the Lick sample compared to literature values. The black line in the upper panels mark the 1:1 correspondence between these parameters. In the lower panel of each diagram we have plotted the corresponding difference or fractional difference of the stellar parameter over the literature value for each giant. The linear best fit to this plot, which takes the full 2D PDFs into account, is represented by the solid blue line, while the dashed blue lines indicate the $95\%$ confidence band of the best fit. Upper left diagram: Comparison of Bayesian stellar radii determined via evolutionary tracks to stellar radii estimated by limb-darkened disk diameters from the CHARM2 catalog \citep{Richichi2005} in conjunction with parallaxes from \cite{vanLeeuwen2007}. We used logarithmic axes since the stellar radii span two orders of magnitude. Upper right diagram: Comparison of Bayesian effective temperature from evolutionary models to spectroscopic effective temperatures by \cite{Hekker2007}. We do not show the error bars here, although they are fully taken into account for the linear fit and in the analysis. The error of the spectroscopic result is always $\unit[84]{K}$ while our median temperature error is around $\unit[33]{K}$. Lower left diagram: As in upper right but for the surface gravity. The error for the spectroscopy for each star is $\unit[0.22]{dex}$ while our median error is $\unit[0.04]{dex}$. Lower right diagram: Comparison of Bayesian stellar masses from evolutionary tracks to the previously determined stellar masses determined by interpolation from evolutionary tracks \citep{Reffert2015}. Green dots indicate that the evolutionary stage determined by both methods are the same while red dots indicate cases where the evolutionary stage differs.}
\label{Fig: all}
\end{figure*}

We compare our results for the effective temperatures to the spectroscopically determined values by \cite{Hekker2007} in Fig.~\ref{Fig: all}, while the statistical results are displayed in Table~\ref{Tab: all}. We find a slight offset of $\unit[-41]{K}$ which is consistent with zero within the errors. \cite{Hekker2007} compared their spectroscopic effective temperatures to other spectroscopic estimates available in the literature and found an offset of $\unit[56]{K}$. We emphasize that such an offset was not found for their metallicities, which were used as an input for our Bayesian method. \cite{Hekker2007} find the largest difference compared to other literature values for the coolest stars, as they slightly overestimate their temperatures. This is in agreement with our difference to their values, as one can see from the upper-right diagram in Fig.~\ref{Fig: all}. In fact our Fig.~\ref{Fig: all} seems to be almost anticorrelated to their Fig.~2. \cite{Hekker2007} argue that the difference is caused by the lower accuracy of their models for low temperatures as well as the increasing number of spectral lines for stars of lower temperatures which leads to line blending. Furthermore, the lines become stronger resulting in a larger dependency on micro turbulence. The reason that the values by \cite{Hekker2007} suffer more from this effect than other spectroscopic determinations might be due to the fact that they used a much smaller number of iron lines than usual to determine the stellar parameters. We conclude that the difference of the effective temperatures is most likely caused by the spectroscopic determinations, as we find exactly the same offset as \cite{Hekker2007}, and therefore our values are fully in line with other literature values.

\subsection{Surface gravity}
\label{spec_logg}

Additionally, we compare the Bayesian surface gravity to the spectroscopic estimates by \cite{Hekker2007}; see the lower-left diagram of Fig.~\ref{Fig: all} and Table~\ref{Tab: all}. We find an average absolute difference of $\langle \text{log}(g)_{\text{trk.}}-\text{log}(g)_{\text{Spec.}}\rangle=\unit[-0.36\pm0.26]{dex}$, and the linear fit parameters show a significant negative slope and offset. \cite{Hekker2007} find an offset of $\unit[0.15]{dex}$ compared to other spectroscopic literature values, which they trace back to their overestimated effective temperatures. This still leaves an average absolute difference of about $ \unit[-0.2]{dex}$ between our values and other spectroscopic estimates. 
 This could mean that we either underestimate stellar masses or overestimate stellar radii. However, we can exclude both possibilities as we do not find evidence for such systematic errors 
     (see Sects.~\ref{stellar_radii} and~\ref{mass_comp}).
     This leads us to the conclusion that it is much more probable that the spectroscopic surface gravities are instead too high. This is a known problem; \cite{Silva2006} observed the same disagreement when they compared their surface gravities determined from stellar isochrones to spectroscopic values. The overestimated spectroscopic gravities can be caused by non-LTE effects on the Fe I abundances \citep{Nissen1997, Lind2012} as well as by the interplay between the stellar parameters that are considered when the abundances are determined \citep{Silva2006}. We refer to \cite{Silva2006} for a detailed discussion of this problem. 
     In Sect.~\ref{surface_gravity}, we verify again that our methodology does not suffer from large systematic biases, using an asteroseismology-based sample for the comparison.

\subsection{Comparison to previous mass and evolutionary stage estimates}
\label{mass_comp}

\begin{table*}[tbp]
\caption{Absolute numbers and fraction of stars in the RGB and HB evolutionary stages, as determined with the various methods.}
\label{Tab: evtable2}
\begin{center}

\begin{tabular}{l l c c c c}
\hline\hline
Method&Evol. St.&$N_{\tx{Stars}}$& $\cap$ Asteroseismology&Recovery rate &Success rate\\
\hline\\ [-1.7ex]
Asteroseismology &RGB&186 ($18\%$)&\ldots &\ldots&\ldots \\
&HB&826 ($82\%$)&\ldots &\ldots&\ldots \\
&Total&1012 ($100\%$)& \ldots &\ldots &\ldots \\
\hline\\ [-1.7ex]
Bayesian inference&RGB&183 ($18\%$)& 115 ($11\%$)&$61.8\%$&$62.8\%$\\
&HB&829 ($82\%$)& 758 ($75\%$)&$91.8\%$&$91.4\%$\\
&Total&1012 ($100\%$)& 873 ($86\%$)&$86.3\%$&\ldots \\
\hline\\ [-1.7ex]
Interpolation&RGB&919 ($91\%$)& 184 ($18\%$)&$98.9\%$&$20.0\%$\\
&HB&93 ($9\%$)& 91 ($9\%$)&$11\%$&$97.8\%$\\
&Total&1012 ($100\%$)& 275 ($27\%$)&$27\%$&\ldots \\
\hline
\end{tabular}
\end{center}
\end{table*}

We compare the masses and evolutionary stages from our Bayesian methodology to previous estimates for the same sample \citep{Reffert2015} based on tri-linear interpolation in the evolutionary tracks of \cite{Girardi2000}. 
Besides using the new evolutionary tracks of \citet{Bressan2013}, our main improvements are the treatment of non-Gaussian errors, which is particularly important for parallaxes, the use of more precise parallax measurements (original Hipparcos Catalogue vs.\ \citeauthor{vanLeeuwen2007} \citeyear{vanLeeuwen2007}), and an improved weighting scheme, which does not a priori favor the RGB over the HB and therefore results in a much more robust estimate of the evolutionary stage. There are no other mass determinations available in the literature for a large enough fraction of our stars to allow for an independent comparison. 

The lower right diagram in Fig.~\ref{Fig: all} shows the Bayesian estimated mass $M_{\text{trk.}}$ over the previously determined mass $M_{\text{p.}}$ from \citet{Reffert2015}, while Table~\ref{Tab: all} shows the statistics for their comparison. We were able to determine stellar parameters for 372 out of 373 stars, while \citet{Reffert2015} derived stellar parameters for 361 stars of the same sample. Furthermore, for 159 stars (44\%), we find a different evolutionary stage compared to the previous estimates. 

The large number of differing evolutionary stages can be explained by the fact that the previous estimates were heavily biased towards the RGB. The reason for this lies in the fact that the interpolation routine will always find solutions for the RGB, as these run continuously through the parameter space occupied by our giants stars, while this is not always the case for the HB tracks, as these start within the occupied parameter space. Observational errors towards bluer colors, for example, will therefore never yield valid interpolation results on the HB. Weighting the number of successful interpolation results of each evolutionary stage by the evolutionary time and by the IMF, as was done in \citet{Reffert2015}, is not enough to account for this effect. The Bayesian method however is able to include the HB models, as it provides the probability of the star belonging to a certain evolutionary track only based on the likelihood and prior probabilities, and therefore does not require the evolutionary tracks to encompass the measured position in the HRD. 

From Table~\ref{Tab: all}, one can see that for the stars for which we find a different evolutionary stage compared to the previous determination, we usually estimate a stellar mass that is smaller. This is expected, as many of these stars are now assigned to be on the HB instead of on the RGB, and the HB evolutionary tracks for a specific mass and metallicity are usually positioned above the corresponding RGB evolutionary tracks in the HRD. The average difference for these stars is $\langle M_{\text{trk.}}-M_{\text{p.}}\rangle=\unit[-0.27\pm0.42]{M_\odot}$, and the linear fit to these stars results in a non-significant slope and intercept. The average difference for the stars that are determined to have the same evolutionary stage by the Bayesian and interpolation method is $\langle M_{\text{trk.}}-M_{\text{p.}}\rangle=\unit[0.00\pm0.48]{M_\odot}$, so the two methods are on average in perfect agreement with each other, although individual cases can of course deviate significantly.

Regarding only the planet hosts in our sample, the overall mass distribution has not changed significantly compared to the results found by \cite{Reffert2015}.

\section{Reliability of our stellar parameter estimates}
\label{Discussion}
\subsection{Evolutionary stages}
\label{evolutionary_stages_discussion}
Our goal in the following is to assess the reliability of the Bayesian methodology regarding the prediction of the post-main sequence evolutionary stage of giant stars, as described in Sect.~\ref{evolution}. For this purpose we used a crossmatched sample (hereafter referred to as the KAG-sample\footnote{KAG-sample stands for Kepler-APOGEE-Gaia Sample}) of 1012 evolved stars in the Kepler field for which asteroseismic evolutionary stages were provided by \cite{Vrard2016}, parallaxes by Gaia DR 1 \citep{Gaia2016}, photometry by 2MASS \citep{2Mass2006} and metallicities by APOGEE \citep{Wilson2010}. While the determination of accurate stellar parameters for this sample of giant stars is currently limited due to the small parallaxes and the relatively large systematic and statistical astrometric errors in the first Gaia data release, it is still possible to compare the predicted evolutionary stages by our Bayesian inference methodology to those based on asteroseismology. This is due to the fact that the determination of the evolutionary stage is most sensitive to the color index, which does not depend on the distance of the star, except for the extinction. We de-reddened this sample by using the extinction in the K magnitude provided in the APOGEE catalog. 

The number of stars in each evolutionary stage, as determined by  various methods (asteroseismology, Bayesian inference and interpolation), for the KAG-sample are shown in Table~\ref{Tab: evtable2}. Furthermore, the table gives the number of stars that are classified in the same way by asteroseismology for the two other methods based on evolutionary tracks. The recovery and success rates in Table~\ref{Tab: evtable2} are defined as the number of such stars classified in the same way by asteroseismology, normalized to the total number of systems classified in that particular evolutionary stage by asteroseismoslogy (recovery rate) and by the number of stars classified to be in that particular evolutionary stage by the method to be tested (success rate), respectively. 

From Table~\ref{Tab: evtable2}, one can see that success and recovery rates for the Bayesian inference method have similar values for both post-main sequence evolutionary stages. Furthermore, the differences between the success and recovery rates for the two evolutionary stages are very small. This shows that our method is not significantly biased towards a certain evolutionary stage. In contrast to this, the evolutionary stages determined by the interpolation method are biased towards the RGB, as can be seen from the table. However, we find that the Bayesian method might slightly favor the HB, as we were not able to recover more than $61.8\%$ of the RGB stars. With smaller observational errors we expect this number to become larger (see also the following section). The total recovery rate of the Bayesian inference method for both evolutionary stages (HB and RGB combined) is around $86.3\%$, which means that for this fraction of stars the Bayesian method provided the same evolutionary stage as asteroseismology. We regard this as a very good result for the determination of the evolutionary stage from the HRD/CMD, in view of the large observational errors of the ABL for this test sample, which is on average a factor of ten larger than the observational errors in our Lick planet search sample. However, we highlight that, in individual cases, even though the probability of finding the correct stage is very high, the determination can still be erroneous. For this reason we usually provide both solutions for the stellar parameters in our catalog. 

\begin{figure}
\resizebox{\hsize}{!}{\includegraphics{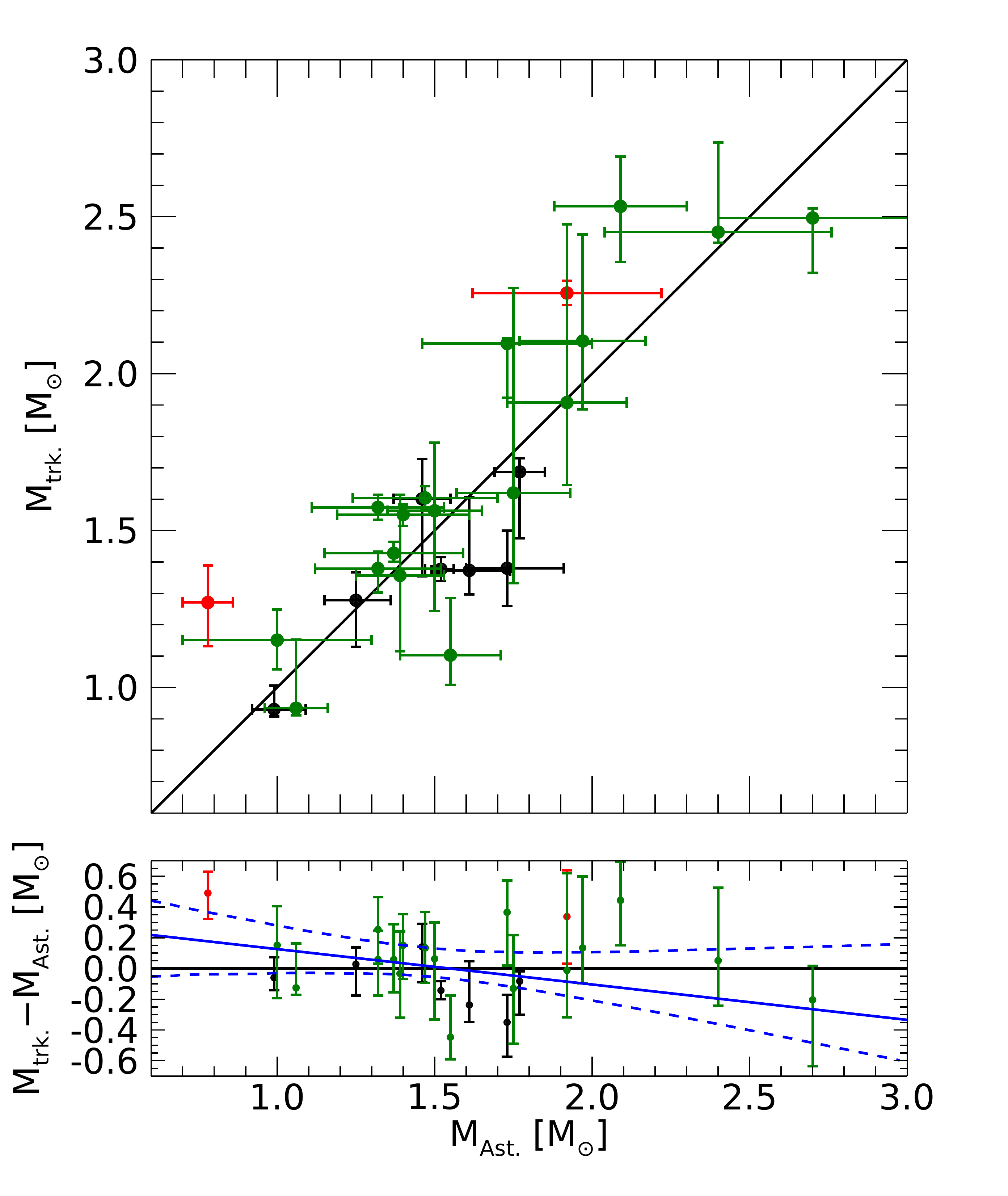}}
\caption{The upper panel shows the stellar masses of 26 giant stars that were determined from evolutionary tracks using our Bayesian methodology compared to asteroseismic reference masses. The black solid line marks the line of equality. The lower panel shows the difference of both masses as a function of the asteroseismic reference mass. Errors of the asteroseismic masses are not plotted for clarity in the lower panel, but are indicated in the upper panel. The blue solid line in the lower panel is a linear fit, where the errors in the asteroseismic masses have been taken fully into account. The blue dashed lines indicate the $95\%$ confidence band.}
\label{Fig: Masses_testsample}
\end{figure}

\subsection{Stellar masses}
\label{Stellar_Mass_discussion}

\begin{table*}[tbp]
\caption{Metallicities, asteroseismic masses $M_\text{Ast.}$ and evolutionary stages $E_\text{Ast.}$ as determined by the reference given in the first column for 26 stars with asteroseismic as well as Bayesian stellar parameters and evolutionary stages available. We also provide Bayesian masses M$_\text{trk.}$ and evolutionary stages $E_\text{B.}$ together with their probability $P_\text{trk.}$.}
\label{Tab: testsample2}
\begin{center}
\begin{tabular}{l r r S[table-format=2.2(3),detect-weight,mode=text] l l l l r}
\hline
\hline
Ref. & Source Identifier &HIP& [{Fe/H}]& $M_\text{Ast.}$ [M$_\odot$]& $E_\text{Ast.}$ & $M_\text{trk.}$ [M$_\odot$]& $E_\text{trk.}$ & $P_\text{trk.}$\\
\hline
1&KIC 03730953 &93687 & -0.07\pm0.03&$1.97\pm0.20$\tablefootmark{a}  & HB &$ 2.104_{-0.218}^{+0.340}$& HB & 0.996\vspace{0.04cm}\\
1&KIC 04044238 &94221 & 0.20\pm0.03&$1.06\pm0.10$\tablefootmark{a}& HB &$ 0.934_{-0.023}^{+0.219} $& HB & 0.824\vspace{0.04cm}\\
1&KIC 05737655 &98269 &  -0.63\pm0.03 &$0.78\pm 0.08$ \tablefootmark{a}& HB &$ 1.271_{-0.139}^{+0.118}$ & RGB & 0.683\vspace{0.04cm}\\
1&KIC 08813946 &95005 & 0.09\pm0.03 &$2.09\pm 0.21 $\tablefootmark{a}& HB &$ 2.533_{-0.177}^{+0.158}$& HB & 0.997\vspace{0.04cm}\\
1&KIC 09705687 &94976 & -0.20\pm0.03 &$1.92\pm 0.19 $\tablefootmark{a}& HB &$ 1.908_{-0.263}^{+0.568}$& HB & 0.981\vspace{0.04cm}\\
1&KIC 10323222 &92885 & 0.04\pm0.03 &$1.55\pm 0.16 $\tablefootmark{a}& RGB &$ 1.103_{-0.095}^{+0.182}$& RGB & 0.999\vspace{0.04cm}\\
1&KIC 10404994 &95687 & -0.06\pm0.03 &$1.50 \pm 0.15 $\tablefootmark{a}& HB &$ 1.563_{-0.319}^{+0.217}$& HB & 0.990\vspace{0.04cm}\\
1&KIC 10716853 &93376 & -0.08\pm0.03 &$1.75\pm 0.18 $\tablefootmark{a}& HB &$ 1.620_{-0.287}^{+0.653}$& HB & 0.990\vspace{0.04cm}\\
1&KIC 12884274 &94896 & 0.11\pm0.03 &$1.39 \pm 0.14 $\tablefootmark{a}& HB &$ 1.357_{-0.241}^{+0.257}$& HB & 0.97\vspace{0.04cm}7\\
2&$\epsilon$ Tau  & 20889 & 0.17\pm0.06{\tablefootmark{b}}&$2.40 \pm 0.36 $&  HB &$ 2.451_{-0.034}^{+0.285}$& HB & 0.993\vspace{0.04cm}\\
2&$\beta$ Gem   & 37826 &0.09\pm0.04{\tablefootmark{b}} &$1.73 \pm 0.27 $& HB &$ 2.096_{-0.173}^{+0.018}$& HB & 0.994\vspace{0.04cm}\\
2&18 Del  &103527 &0.07\pm0.04{\tablefootmark{c}} &$1.92 \pm 0.30 $& HB &$ 2.257_{-0.039}^{+0.039}$& RGB & 0.994\vspace{0.04cm}\\
2&$\gamma$ Cep  &116727 &0.13\pm0.06 &$ 1.32 \pm 0.20 $& RGB &$ 1.379_{-0.077}^{+0.054}$& RGB & 1.000\vspace{0.04cm}\\
2&HD 5608  &4552 &0.12\pm0.03 &$1.32 \pm 0.21 $&  RGB &$ 1.574\pm0.040$& RGB & 1.000\vspace{0.04cm}\\
2&$\kappa$ CrB  &77655 &0.13\pm0.03 &$1.40 \pm 0.21 $& RGB &$ 1.551_{-0.036}^{+0.032}$& RGB & 1.000\vspace{0.04cm}\\
2&6 Lyn  &31039 &-0.13\pm0.02 &$1.37 \pm 0.22 $& RGB &$ 1.428_{-0.027}^{+0.036}$& RGB & 1.000\vspace{0.04cm}\\
2&HD 210702  &109577 &0.04\pm0.03  &$1.47 \pm 0.23 $& RGB &$ 1.604_{-0.034}^{+0.038}$& RGB & 1.000\vspace{0.04cm}\\
3&HD 145428  &79364 & -0.32\pm0.12&$0.99_{-0.07}^{+0.10} $& \ldots &$ 0.930_{-0.022}^{+0.076}$& RGB & 1.000\vspace{0.04cm}\\
3&HD 4313  &3574 & 0.05\pm0.1 &$1.61_{-0.12}^{+0.13} $& \ldots &$ 1.373_{-0.076}^{+0.234}$ & RGB & 1.000\vspace{0.04cm}\\
3&HD 181342  &95124 &0.15\pm0.1 &$1.73_{-0.13}^{+0.18} $& \ldots &$ 1.380\pm0.120 $& RGB & 1.000\vspace{0.04cm}\\
3&HD 5319  &4297 &0.02\pm0.1 &$1.25_{-0.10}^{+0.11} $& \ldots &$ 1.278_{-0.149}^{+0.089}$& RGB & 1.000\vspace{0.04cm}\\
3&HD 185351  & 96459 &0.00\pm0.1{\tablefootmark{b}} &$1.77 \pm 0.08  $& \ldots &$ 1.687_{-0.221}^{+0.043}$& RGB & 1.000\vspace{0.04cm}\\
3&HD 212771  &110813 &-0.1\pm0.12 &$1.46 \pm 0.09 $& \ldots &$ 1.601_{-0.247}^{+0.127}$& RGB & 1.000\vspace{0.04cm}\\
3&HD 106270  &59625 &0.06\pm0.1 &$1.52_{-0.05}^{+0.04} $& \ldots &$ 1.377_{-0.037}^{+0.038}$& RGB & 1.000\vspace{0.04cm}\\
4&$\gamma$ Piscium  & 114971 &-0.54\pm0.10 {\tablefootmark{b}}&$1.00\pm0.30${\tablefootmark{d}} & HB{\tablefootmark{e}} &$ 1.151_{-0.093}^{+0.097}$& HB & 0.907\vspace{0.04cm}\\
4&$\Theta^1$ Tauri  & 20885 &0.08\pm0.10 {\tablefootmark{b}}&$2.70\pm 0.30 $& HB{\tablefootmark{e}} &$ 2.496_{-0.175}^{+0.031} $& HB & 0.997\vspace{0.04cm}\\
\hline
\end{tabular}

\tablebib{
(1)~\citet{Takeda2015}; (2) \citet{Stello2017}; (3) \citet{North2017} ; (4) \citet{Beck2015}
}
\tablefoot{
\tablefoottext{a}{no confidence interval provided; we estimated a $10\%$ error from Fig.~10 in \cite{Takeda2015}}
\tablefoottext{b}{metallicity from \cite{Hekker2007}}
\tablefoottext{c}{metallicity from \cite{Maldonado2013}}
\tablefoottext{d}{no confidence interval provided; we assumed the same as for HIP 20885}
\tablefoottext{e}{no mixed modes available, \cite{Beck2015} estimated the evolutionary stage from the frequency of
the maximum oscillation power excess}
}
\end{center}
\end{table*}
In order to determine the reliability of our methodology to estimate the stellar mass of the star, we created a test sample consisting of stars that have photometry and parallaxes available by Hipparcos (so that we can apply the Bayesian method) as well as asteroseismically determined reference masses. For all cases, we used the metallicities that were provided by the references, with the exception of stars that are also in our main sample (HIP~20885, HIP~20889, HIP~31592, HIP~96459 and HIP~114971), for which we used the spectroscopically determined values by \citet{Hekker2007}. Our test sample consists of 26 stars. Nine of these stars were investigated by \cite{Takeda2015}, eight stars by \cite{Stello2017}, seven stars by \cite{North2017}, and two stars by \cite{Beck2015}. The input metallicities and their references, as well as the asteroseismic values are provided in Table~\ref{Tab: testsample2}. \cite{Takeda2015} did not provide a confidence interval for their asteroseismic masses; based on their Fig.~10 we estimated an error of $10\%$. \cite{Beck2015} also did not provide an error for one of their stars; we used a conservative value of $\unit[0.3]{M_\odot}$. With the exception of the stars from \cite{North2017} we also have the asteroseismic evolutionary stages available and therefore compare them to the derived evolutionary stages with the Bayesian method. 

For the 19 stars with available asteroseismic evolutionary stages, the Bayesian method provides the same evolutionary stage for 17 stars, in the sense that the probability of the evolutionary stage is higher than $50\%$. This is again a total recovery rate of $\sim\!\!89\%$. The probability of being on the RGB for the seven stars that are missing an evolutionary stage from asteroseismology was $100\%$ according to our method. By inspecting these stars in the HRD, we find that it is very unlikely that these stars could be attributed to the HB. Therefore, we include these stars in our comparison of the stellar masses.

Figure~\ref{Fig: Masses_testsample} shows the comparison of Bayesian masses with asteroseismic masses as well as the differences of both as a function of the asteroseismic mass. The green dots are stars for which the evolutionary stage determination agrees between both methods, while the red dots indicate the cases where the Bayesian method found a different evolutionary stage than found by asteroseismology. The black dots are stars for which no asteroseismic evolutionary stage was available. The resulting fit parameters of the best fit (blue solid line) for the 24 stars that have probably a correct assigned evolutionary stage by our method are provided in Table~\ref{Tab: all}. 
For these 24 stars we find an average mass difference of only $\langle\Delta M\rangle=\langle M_{\text{trk.}}-M_{\text{Ast.}}\rangle=\unit[0.01\pm0.20]{M_\odot}$. Including the two stars with different evolutionary stages, we determine $\langle\Delta M\rangle=\langle M_{\text{trk.}}-M_{\text{Ast.}}\rangle=\unit[0.04\pm0.22]{M_\odot}$, which is still not significantly different from zero within the error. 
We point out that for one star (HIP 98269), with differing evolutionary stage and asteroseismic mass $M_{\text{Ast.}}=\unit[0.78]{M_\odot}$, the alternate solution was just slightly less probable and would have led to much better agreement of the masses ($\unit[0.81^{+0.11}_{-0.02}]{M_\odot}$ in the HB case, compared to $\unit[1.271_{-0.139}^{+0.118}]{M_\odot}$ for the slightly more probable RGB case). 

Regarding the linear fits we used two approaches. We fitted for the differences between both masses, taking both the error of the difference and the error of the reference value (asteroseismic mass) into account as done throughout this work. However, we also determined the linear fit parameters neglecting the errors of the asteroseismic mass determinations as often seen in the literature. This was done because some errors of the asteroseismic masses were not provided and therefore needed to be estimated.

Regarding the linear fits that take only the errors of the differences into account, we find no significant slope or offset for the sample of 24 stars. However, taking into account the errors of the asteroseismic masses leads to a more significant positive offset ($\unit[0.36^{+0.17}_{-0.21}]{M_\odot}$) and negative slope ($\unit[-0.21^{+0.14}_{-0.11}]{}$) for the linear fit. This result would indicate that we slightly overestimate stellar masses for lower-mass stars and underestimate stellar masses for stars with masses higher than $\unit[\sim\!\!2]{M_\odot}$ compared to asteroseismology. However, it must be mentioned that the errors of the asteroseismic masses for a significant fraction of our sample were estimated by us, as the references did not provide an error estimate. %Furthermore, regarding the test sample of stars for which we determined the linear fits, one can see that most stars are in the region between $\unit[1.2]{M_\odot}$ to $\unit[1.7]{M_\odot}$ where one can see a small negative linear trend. Furthermore, the stars with the lowest and highest asteroseismic masses in the test sample follow this trend. The lack of more stars with lower and higher stellar masses limits the validity of the fitting results.}
The slopes and offsets of the linear fits that take all errors into account are within the 2$\sigma$ range of zero for the 24 stars. %Due to this and the fact that the absolute average difference is very small as well as the significance of the linear fit parameters when we neglect the error of the (in some cases estimated) asteroseismic masses,
Therefore, we do not find evidence for strong systematic errors between stellar masses from evolutionary models and asteroseismic masses. In particular, we do not find any evidence that stellar masses of so-called retired A stars determined from evolutionary models are significantly overestimated, in contrast to \cite{Lloyd2011}, \cite{SchlaufmanWin2013} and \cite{Sousa2015}. Regarding only the subsample that is taken from \cite{Stello2017} we recognize that all our stellar masses are slightly larger than the asteroseismic masses. However, the difference is not very significant and needs further investigation with a larger sample of stars in the future. %We conclude that the stellar masses we determined for our Lick giants do not suffer from a systematic overestimation or any other bias.

\subsection{Surface gravity}
\label{surface_gravity}

\begin{figure}
\resizebox{\hsize}{!}{\includegraphics{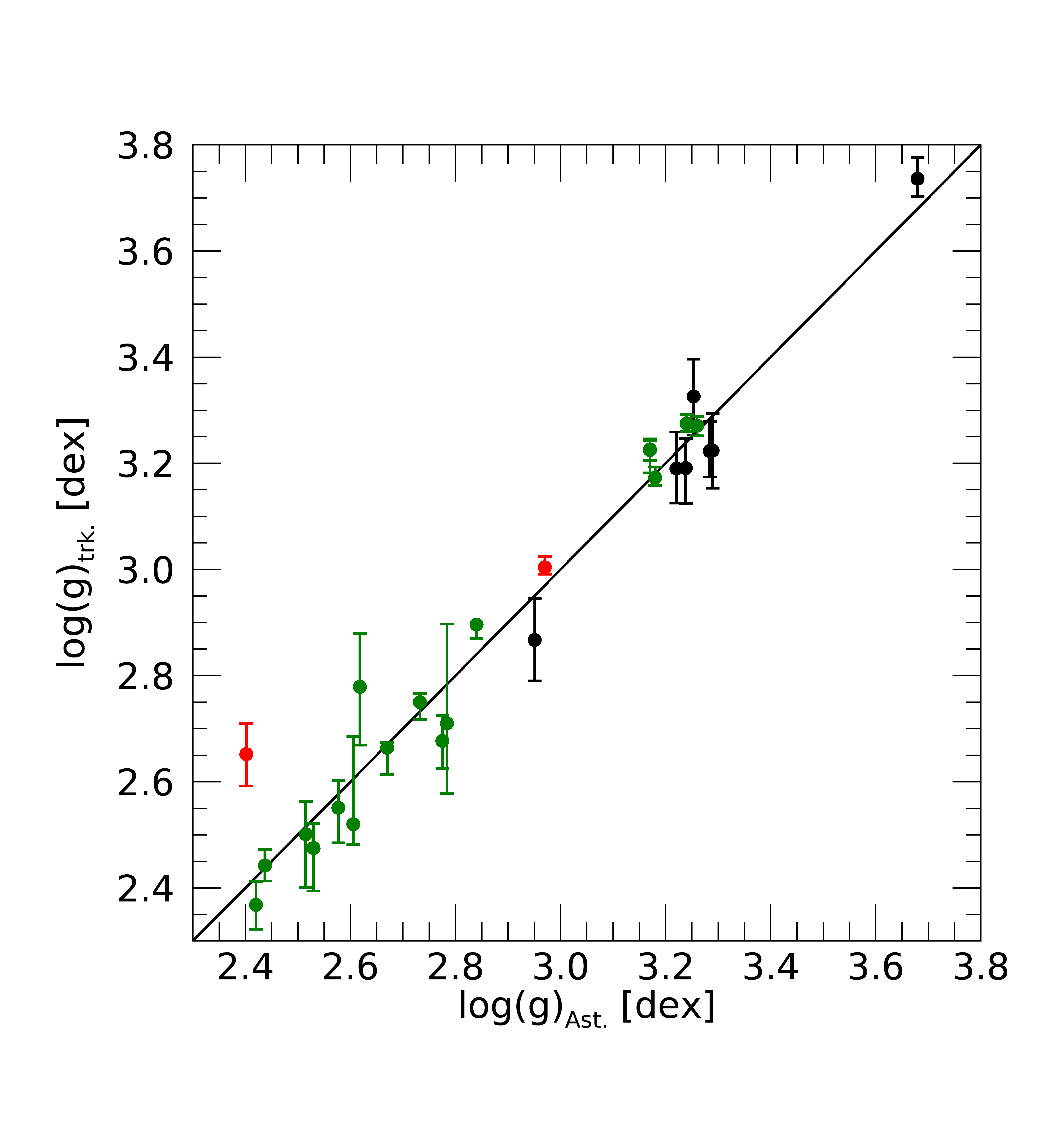}}
\caption{Comparison of the surface gravities derived from evolutionary tracks to asteroseismic surface gravities. Colors and the black line have the same meaning as in Fig.~\ref{Fig: Masses_testsample}.}
\label{Fig: finalplot_logg_astro}
\end{figure}

We checked the reliability of our estimates of the surface gravity by comparing the values in our test sample of 26 stars to the surface gravities that are derived by using the asteroseismic mass and radius. We determine an absolute average difference of only $\langle \text{log}(g)_{\text{trk.}}-\text{log}(g)_{\text{Ast.}}\rangle=\unit[0.004\pm0.077]{dex}$. This confirms our earlier conclusion (see Sect. \ref{spec_logg}) that the spectroscopic surface gravities from \cite{Hekker2007} are indeed too high, and that our estimates of the surface gravity do not suffer from a large systematic bias. Figure~\ref{Fig: finalplot_logg_astro} shows the comparison of Bayesian surface gravities to surface gravities determined from asteroseismic radii and masses.

\section{Summary and conclusions}
\label{Summary}

We used a Bayesian methodology that includes the initial mass function and evolutionary times of the stars as a prior to estimate stellar parameters  by comparing the position of the stars in the Hertzsprung-Russel diagram to grids of evolutionary tracks.
We have estimated the evolutionary stages and associated stellar parameters of 372 of the 373 giant stars in our sample. We find that 70 stars (19\%) are most probably on the RGB, while the remaining 302 (81\%) are probably on the HB. We compared stellar radii of a sub-sample of 86 stars to stellar radii based on measured limb-darkened disk diameters from CHARM2 \citep{Richichi2005} and Hipparcos parallaxes by \cite{vanLeeuwen2007}. We found no significant systematic differences between the stellar radii determined with the two different methods. Additionally, we compared the stellar effective temperatures and surface gravities of  97\% of the stars in our sample to spectroscopic estimates. While we found small systematic differences of our estimates compared to the spectroscopic measurements, we were able to attribute these differences to the spectroscopic estimates. 

Recently several studies stated that stellar masses determined from a grid of evolutionary models can be overestimated significantly \citep{Lloyd2011,Lloyd2013,SchlaufmanWin2013,Sousa2015}. We tested the reliability of our methodology to derive stellar masses by comparing mass values derived with our Bayesian method to asteroseismic reference masses for stars in a small test sample. We found no large differences regarding the average absolute differences. A linear fit to the individual differences as a function of the asteroseismic masses, taking the uncertainties of Bayesian and the asteroseismic masses fully into account, resulted in a slightly negative slope and positive offset. This would mean that our masses for the more massive stars in our sample are slightly underestimated compared to asteroseismology, in contrast to the implications of \cite{Lloyd2011} and \cite{SchlaufmanWin2013}. However, the linear fit parameters are not very significant and are most probably largely influenced by the small number of stars at the boundaries of the occupied parameter space. By comparing our estimated stellar masses to asteroseismic reference masses, we do not find evidence that stellar masses from evolutionary models are significantly overestimated by the order of magnitude that was stated by \cite{Lloyd2011}, \cite{SchlaufmanWin2013} and \cite{Sousa2015}. However, we notice that for some stars, a large spread of stellar masses determined from isochrones or evolutionary models is available in the literature. There are many reasons for this large spread.

Systematic differences can be caused by the adoption of different stellar models, as these can differ in their positions in the Hertzsprung-Russel diagram as well as their input physics \citep{Cassisi2012, North2017}. Another source of uncertainty are the bolometric corrections that are applied to stellar models if photometry is included to determine the position of the stars in the color magnitude diagram. However, these uncertainties are probably smaller than the systematic uncertainties that are included if spectroscopic surface gravities and effective temperatures are used instead of photometry. 
In order to be able to recover a good estimate of the stellar parameters from evolutionary models, it is essential that the uncertainties of the input parameters are not neglected or underestimated. Furthermore, for the determination of stellar parameters from evolutionary models, one should include either a prior or some sort of weighting so that unrealistic or improbable solutions can be ruled out a priori. 

Due to the degeneracy of the post-main sequence models in the Hertzprung-Russel diagram, it is also important to either estimate the evolutionary stages of giant stars or provide the stellar parameters of both stages, as the adoption of one evolutionary stage over another can lead to significant differences in the stellar parameters. This is often neglected or not properly taken into account. For 159 stars of our sample, we find a different evolutionary stage with the Bayesian method from this work compared to previous estimates that were less reliable regarding the determination of the evolutionary stage. For these 159 stars, we find on average a mass that is $\sim\!\!\unit[0.3]{M_\odot}$ smaller. 

Regarding the planet hosting stellar systems in our sample we find that for the 16 giants with planets, 15 have a higher probability of being on the horizontal branch. The planet hosts are HIP~75458 \citep{Frink2002}, HIP~37826 \citep{Reffert2006}, HIP~88048 \citep{Quirrenbach2011}, HIP 34693 and HIP~114855 \citep{Mitchell2013}, HIP~5364 \citep{Trifonov2014}, HIP~36616 \citep{Ortiz2016}, HIP~20889 \citep{Sato2007} and HIP~60202 \citep{Liu2008} as well as seven systems whose publications are in preparation. We excluded HIP~21421 (Aldebaran), a star that is probably on the RGB and has a published planet by \cite{Hatzes2015}. However, the existence of this planet is very uncertain \citep{Reichert, Hatzes2018}. Furthermore, we do not find significant differences for the adopted stellar masses of the planet-hosting stars that were used for the analysis of \cite{Reffert2015} regarding planet occurrence rates around giant stars.

After work on this paper was concluded, the Gaia DR2 parallaxes became
available \citep{Gaia2018}. Only 217 of the 373 stars in our Lick sample have a parallax measurement in Gaia DR2. While the astrometric accuracy of
Hipparcos is higher for brighter stars, the opposite is true for Gaia,
meaning that for 44 of our stars (roughly those brighter than $V\sim4.5$~-~$\unit[5.0]{mag}$) the Hipparcos parallaxes are actually more accurate than the ones from Gaia DR2. That leaves 173 stars (46\% of the stars in our sample)
for which the Gaia DR2 parallaxes are formally more accurate. However,  the improvement
for these stars is not nearly as dramatic as for fainter stars. The results of this work are therefore not significantly affected by using the Hipparcos parallaxes by \cite{vanLeeuwen2007} instead of the parallaxes from Gaia DR2.

Overall, we come to the conclusion that the estimation of stellar parameters and evolutionary stages for giant stars derived from stellar evolutionary tracks by using spectroscopic, photometric, and astrometric observables is valid, as long as the uncertainties of the input parameters, as well as physical prior knowledge, is carefully taken into account.

\begin{acknowledgements} We would like to thank G.~Worthey, L.~Girardi and L.~Lindegren for helpful and useful email discussions. A special thanks goes to A.~Bressan for his very valueable input on various occasions. This work was supported by the DFG Research Unit FOR2544 'Blue Planets around Red Stars', project no.~RE 2694/4-1. This research has made use of the VizieR catalog access tool, CDS, Strasbourg, France. The original description of the VizieR
service was published in A\&AS 143, 23.
This research has made use of the TOPCAT (http://www.starlink.ac.uk/topcat/) and STILTS (http://www.starlink.ac.uk/stilts/) software, provided by  
Mark Taylor of Bristol University, England. \end{acknowledgements}

\bibpunct{(}{)}{;}{a}{}{,} % to follow the A&A style
% for the bibliography, at the end
\bibliographystyle{aa} % style aa.bst
%\bibliography{references.bib}
\bibliography{references.bib}

\begin{thebibliography}{77}
\expandafter\ifx\csname natexlab\endcsname\relax\def\natexlab#1{#1}\fi

\bibitem[{{Arenou} \& {Luri}(1999)}]{Arenou1999}
{Arenou}, F. \& {Luri}, X. 1999, in Astronomical Society of the Pacific
  Conference Series, Vol. 167, Harmonizing Cosmic Distance Scales in a
  Post-HIPPARCOS Era, ed. D.~{Egret} \& A.~{Heck}, 13--32

\bibitem[{{Astraatmadja} \& {Bailer-Jones}(2016)}]{bailer2016}
{Astraatmadja}, T.~L. \& {Bailer-Jones}, C.~A.~L. 2016, ApJ, 832, 137

\bibitem[{{Bailer-Jones}(2015)}]{bailer2015}
{Bailer-Jones}, C.~A.~L. 2015, PASP, 127, 994

\bibitem[{{Beck} {et~al.}(2015){Beck}, {Kambe}, {Hillen}, {Corsaro}, {Van
  Winckel}, {Moravveji}, {De Ridder}, {Bloemen}, {Saesen}, {Mathias},
  {Degroote}, {Kallinger}, {Verhoelst}, {Ando}, {Carrier}, {Acke}, {Oreiro},
  {Miglio}, {Eggenberger}, {Sato}, {Zwintz}, {P{\'a}pics}, {Marcos-Arenal},
  {Sans Fuentes}, {Schmid}, {Waelkens}, {{\O}stensen}, {Matthews}, {Yoshida},
  {Izumiura}, {Koyano}, {Nagayama}, {Shimizu}, {Okada}, {Okita}, {Sakamoto},
  {Yamamuro}, \& {Aerts}}]{Beck2015}
{Beck}, P.~G., {Kambe}, E., {Hillen}, M., {et~al.} 2015, A\&A, 573, A138

\bibitem[{{Bressan} {et~al.}(2013){Bressan}, {Marigo}, {Girardi}, {Nanni}, \&
  {Rubele}}]{Bressan2013}
{Bressan}, A., {Marigo}, P., {Girardi}, L., {Nanni}, A., \& {Rubele}, S. 2013,
  in European Physical Journal Web of Conferences, Vol.~43, European Physical
  Journal Web of Conferences, 03001

\bibitem[{{Bressan} {et~al.}(2012){Bressan}, {Marigo}, {Girardi}, {Salasnich},
  {Dal Cero}, {Rubele}, \& {Nanni}}]{Bressan2012}
{Bressan}, A., {Marigo}, P., {Girardi}, L., {et~al.} 2012, MNRAS, 427, 127

\bibitem[{{Cassisi}(2012)}]{Cassisi2012}
{Cassisi}, S. 2012, Astrophys. Space Sci. Proc., 26, 57

\bibitem[{{Currie}(2009)}]{Currie2009}
{Currie}, T. 2009, ApJL, 694, L171

\bibitem[{{da Silva} {et~al.}(2006){da Silva}, {Girardi}, {Pasquini},
  {Setiawan}, {von der L{\"u}he}, {de Medeiros}, {Hatzes}, {D{\"o}llinger}, \&
  {Weiss}}]{Silva2006}
{da Silva}, L., {Girardi}, L., {Pasquini}, L., {et~al.} 2006, A\&A, 458, 609

\bibitem[{Epanechnikov(1969)}]{Ep1969}
Epanechnikov, V.~A. 1969, Theory Probab. Appl., 14:1, 153–158

\bibitem[{ESA(1997)}]{Hipparcos1997}
ESA. 1997, The Hipparcos and Tycho Catalogues, ESA SP-1200

\bibitem[{{Feuillet} {et~al.}(2016){Feuillet}, {Bovy}, {Holtzman}, {Girardi},
  {MacDonald}, {Majewski}, \& {Nidever}}]{feuillet2016}
{Feuillet}, D.~K., {Bovy}, J., {Holtzman}, J., {et~al.} 2016, ApJ, 817, 40

\bibitem[{{Franchini} {et~al.}(2004){Franchini}, {Morossi}, {Di Marcantonio},
  {Malagnini}, {Chavez}, \& {Rodr{\'{\i}}guez-Merino}}]{Franchini2004}
{Franchini}, M., {Morossi}, C., {Di Marcantonio}, P., {et~al.} 2004, ApJ, 613,
  312

\bibitem[{{Frink} {et~al.}(2002){Frink}, {Mitchell}, {Quirrenbach}, {Fischer},
  {Marcy}, \& {Butler}}]{Frink2002}
{Frink}, S., {Mitchell}, D.~S., {Quirrenbach}, A., {et~al.} 2002, ApJ, 576, 478

\bibitem[{{Frink} {et~al.}(2001){Frink}, {Quirrenbach}, {Fischer}, {R{\"o}ser},
  \& {Schilbach}}]{Frink2001}
{Frink}, S., {Quirrenbach}, A., {Fischer}, D., {R{\"o}ser}, S., \& {Schilbach},
  E. 2001, PASP, 113, 173

\bibitem[{{Gaia Collaboration} {et~al.}(2018){Gaia Collaboration}, {Brown},
  {Vallenari}, {Prusti}, {de Bruijne}, {Babusiaux}, \&
  {Bailer-Jones}}]{Gaia2018}
{Gaia Collaboration}, {Brown}, A.~G.~A., {Vallenari}, A., {et~al.} 2018, ArXiv
  e-prints

\bibitem[{{Gaia Collaboration} {et~al.}(2016){Gaia Collaboration}, {Brown},
  {Vallenari}, {Prusti}, {de Bruijne}, {Mignard}, {Drimmel}, {Babusiaux},
  {Bailer-Jones}, {Bastian}, \& et~al.}]{Gaia2016}
{Gaia Collaboration}, {Brown}, A.~G.~A., {Vallenari}, A., {et~al.} 2016, A\&A,
  595, A2

\bibitem[{{Ghezzi} \& {Johnson}(2015)}]{Ghezzi2015}
{Ghezzi}, L. \& {Johnson}, J.~A. 2015, ApJ, 812, 96

\bibitem[{{Girardi} {et~al.}(2000){Girardi}, {Bressan}, {Bertelli}, \&
  {Chiosi}}]{Girardi2000}
{Girardi}, L., {Bressan}, A., {Bertelli}, G., \& {Chiosi}, C. 2000, A\&AS, 141,
  371

\bibitem[{{Girardi} {et~al.}(2013){Girardi}, {Marigo}, {Bressan}, \&
  {Rosenfield}}]{Girardi2013}
{Girardi}, L., {Marigo}, P., {Bressan}, A., \& {Rosenfield}, P. 2013, ApJ, 777,
  142

\bibitem[{{Giridhar} {et~al.}(1997){Giridhar}, {Ferro}, \&
  {Parrao}}]{Giridhar1997}
{Giridhar}, S., {Ferro}, A., \& {Parrao}, L. 1997, PASP, 109, 1077

\bibitem[{{Hansen} \& {Kjaergaard}(1971)}]{Hansen1971}
{Hansen}, L. \& {Kjaergaard}, P. 1971, A\&A, 15, 123

\bibitem[{{Hatzes} {et~al.}(2015){Hatzes}, {Cochran}, {Endl}, {Guenther},
  {MacQueen}, {Hartmann}, {Zechmeister}, {Han}, {Lee}, {Walker}, {Yang},
  {Larson}, {Kim}, {Mkrtichian}, {D{\"o}llinger}, {Simon}, \&
  {Girardi}}]{Hatzes2015}
{Hatzes}, A.~P., {Cochran}, W.~D., {Endl}, M., {et~al.} 2015, A\&A, 580, A31

\bibitem[{{Hatzes} {et~al.}(2018){Hatzes}, {Endl}, {Cochran}, {MacQueen},
  {Han}, {Lee}, {Kim}, {Mkrtichian}, {D{\"o}llinger}, {Hartmann},
  {Karjalainen}, \& {Dreizler}}]{Hatzes2018}
{Hatzes}, A.~P., {Endl}, M., {Cochran}, W.~D., {et~al.} 2018, \aj, 155, 120

\bibitem[{{Hatzes} {et~al.}(2004){Hatzes}, {Setiawan}, {Pasquini}, \& {da
  Silva}}]{Hatzes2004}
{Hatzes}, A.~P., {Setiawan}, J., {Pasquini}, L., \& {da Silva}, L. 2004, in ESA
  Special Publication, Vol. 538, Stellar Structure and Habitable Planet
  Finding, ed. F.~{Favata}, S.~{Aigrain}, \& A.~{Wilson}, 87--92

\bibitem[{{Hayes} {et~al.}(1985){Hayes}, {Pasinetti}, \& {Philip}}]{Hayes1985}
{Hayes}, D.~S., {Pasinetti}, L.~E., \& {Philip}, A.~G.~D., eds. 1985, IAU
  Symposium, Vol. 111, {Calibration of fundamental stellar quantities;
  Proceedings of the Symposium, Como, Italy, May 24-29, 1984}

\bibitem[{{Hekker} \& {Mel{\'e}ndez}(2007)}]{Hekker2007}
{Hekker}, S. \& {Mel{\'e}ndez}, J. 2007, A\&A, 475, 1003

\bibitem[{{Hekker} {et~al.}(2006){Hekker}, {Reffert}, {Quirrenbach},
  {Mitchell}, {Fischer}, {Marcy}, \& {Butler}}]{Hekker2006_2}
{Hekker}, S., {Reffert}, S., {Quirrenbach}, A., {et~al.} 2006, A\&A, 454, 943

\bibitem[{{Hekker} {et~al.}(2008){Hekker}, {Snellen}, {Aerts}, {Quirrenbach},
  {Reffert}, \& {Mitchell}}]{Hekker2008}
{Hekker}, S., {Snellen}, I.~A.~G., {Aerts}, C., {et~al.} 2008, A\&A, 480, 215

\bibitem[{{Hogg}(2008)}]{Hogg2008}
{Hogg}, D.~W. 2008, ArXiv e-prints

\bibitem[{{Hrudkov{\'a}} {et~al.}(2017){Hrudkov{\'a}}, {Hatzes}, {Karjalainen},
  {Lehmann}, {Hekker}, {Hartmann}, {Tkachenko}, {Prins}, {Van Winckel}, {De
  Nutte}, {Dumortier}, {Fr{\'e}mat}, {Hensberge}, {Jorissen}, {Lampens},
  {Laverick}, {Lombaert}, {P{\'a}pics}, {Raskin}, {S{\'o}dor}, {Thoul}, {Van
  Eck}, \& {Waelkens}}]{Hrudkova2017}
{Hrudkov{\'a}}, M., {Hatzes}, A., {Karjalainen}, R., {et~al.} 2017, MNRAS, 464,
  1018

\bibitem[{John(1982)}]{John1982}
John, S. 1982, Commun. in Statistics - Theory and Methods, 11, 879

\bibitem[{{J{\o}rgensen} \& {Lindegren}(2005)}]{Jorgendsen2005}
{J{\o}rgensen}, B.~R. \& {Lindegren}, L. 2005, A\&A, 436, 127

\bibitem[{{Kunitomo} {et~al.}(2011){Kunitomo}, {Ikoma}, {Sato}, {Katsuta}, \&
  {Ida}}]{Kunitomo2011}
{Kunitomo}, M., {Ikoma}, M., {Sato}, B., {Katsuta}, Y., \& {Ida}, S. 2011, ApJ,
  737, 66

\bibitem[{{Lind} {et~al.}(2012){Lind}, {Bergemann}, \& {Asplund}}]{Lind2012}
{Lind}, K., {Bergemann}, M., \& {Asplund}, M. 2012, MNRAS, 427, 50

\bibitem[{Liu {et~al.}(2008)Liu, Sato, Zhao, Noguchi, Wang, Kambe, Ando,
  Izumiura, Chen, Okada, Toyota, Omiya, Masuda, Takeda, Murata, Itoh, Yoshida,
  Kokubo, \& Ida}]{Liu2008}
Liu, Y.-J., Sato, B., Zhao, G., {et~al.} 2008, ApJ, 672, 553

\bibitem[{{Lloyd}(2011)}]{Lloyd2011}
{Lloyd}, J.~P. 2011, ApJL, 739, L49

\bibitem[{{Lloyd}(2013)}]{Lloyd2013}
{Lloyd}, J.~P. 2013, ApJL, 774, L2

\bibitem[{{Luque} {et~al.}(in prep.){Luque}, {Reffert}, \&
  {Quirrenbach}}]{Luque2018}
{Luque}, R., {Reffert}, S., \& {Quirrenbach}, A. in prep., A\&A

\bibitem[{{Lutz} \& {Kelker}(1973)}]{Lutz1973}
{Lutz}, T.~E. \& {Kelker}, D.~H. 1973, PASP, 85, 573

\bibitem[{{Maldonado} {et~al.}(2013){Maldonado}, {Villaver}, \&
  {Eiroa}}]{Maldonado2013}
{Maldonado}, J., {Villaver}, E., \& {Eiroa}, C. 2013, A\&A, 554, A84

\bibitem[{{Mayor} \& {Queloz}(1995)}]{Mayor1995}
{Mayor}, M. \& {Queloz}, D. 1995, Nature, 378, 355

\bibitem[{{Miglio} {et~al.}(2012){Miglio}, {Brogaard}, {Stello}, {Chaplin},
  {D'Antona}, {Montalb{\'a}n}, {Basu}, {Bressan}, {Grundahl}, {Pinsonneault},
  {Serenelli}, {Elsworth}, {Hekker}, {Kallinger}, {Mosser}, {Ventura},
  {Bonanno}, {Noels}, {Silva Aguirre}, {Szabo}, {Li}, {McCauliff}, {Middour},
  \& {Kjeldsen}}]{Miglio2012}
{Miglio}, A., {Brogaard}, K., {Stello}, D., {et~al.} 2012, MNRAS, 419, 2077

\bibitem[{{Mitchell} {et~al.}(2013){Mitchell}, {Reffert}, {Trifonov},
  {Quirrenbach}, \& {Fischer}}]{Mitchell2013}
{Mitchell}, D.~S., {Reffert}, S., {Trifonov}, T., {Quirrenbach}, A., \&
  {Fischer}, D.~A. 2013, A\&A, 555, A87

\bibitem[{{Ness} {et~al.}(2016){Ness}, {Hogg}, {Rix}, {Martig}, {Pinsonneault},
  \& {Ho}}]{Ness2016}
{Ness}, M., {Hogg}, D.~W., {Rix}, H.-W., {et~al.} 2016, ApJ, 823, 114

\bibitem[{{Nissen} {et~al.}(1997){Nissen}, {Hoeg}, \& {Schuster}}]{Nissen1997}
{Nissen}, P.~E., {Hoeg}, E., \& {Schuster}, W.~J. 1997, in ESA Special
  Publication, Vol. 402, Hipparcos - Venice '97, ed. R.~M. {Bonnet},
  E.~{H{\o}g}, P.~L. {Bernacca}, L.~{Emiliani}, A.~{Blaauw}, C.~{Turon},
  J.~{Kovalevsky}, L.~{Lindegren}, H.~{Hassan}, M.~{Bouffard}, B.~{Strim},
  D.~{Heger}, M.~A.~C. {Perryman}, \& L.~{Woltjer}, 225--230

\bibitem[{{North} {et~al.}(2017){North}, {Campante}, {Miglio}, {Davies},
  {Grunblatt}, {Huber}, {Kuszlewicz}, {Lund}, {Cooke}, \&
  {Chaplin}}]{North2017}
{North}, T.~S.~H., {Campante}, T.~L., {Miglio}, A., {et~al.} 2017, MNRAS, 472,
  1866

\bibitem[{{Ortiz} {et~al.}(2016){Ortiz}, {Reffert}, {Trifonov}, {Quirrenbach},
  {Mitchell}, {Nowak}, {Buenzli}, {Zimmerman}, {Bonnefoy}, {Skemer},
  {Defr{\`e}re}, {Lee}, {Fischer}, \& {Hinz}}]{Ortiz2016}
{Ortiz}, M., {Reffert}, S., {Trifonov}, T., {et~al.} 2016, A\&A, 595, A55

\bibitem[{{Perryman} {et~al.}(1997){Perryman}, {Lindegren}, {Kovalevsky},
  {Hoeg}, {Bastian}, {Bernacca}, {Cr{\'e}z{\'e}}, {Donati}, {Grenon},
  {Grewing}, {van Leeuwen}, {van der Marel}, {Mignard}, {Murray}, {Le Poole},
  {Schrijver}, {Turon}, {Arenou}, {Froeschl{\'e}}, \&
  {Petersen}}]{Perryman1997a}
{Perryman}, M., {Lindegren}, L., {Kovalevsky}, J., {et~al.} 1997, A\&A, 323,
  L49

\bibitem[{{Quirrenbach} {et~al.}(2011){Quirrenbach}, {Reffert}, \&
  {Bergmann}}]{Quirrenbach2011}
{Quirrenbach}, A., {Reffert}, S., \& {Bergmann}, C. 2011, in American Institute
  of Physics Conference Series, Vol. 1331, American Institute of Physics
  Conference Series, ed. S.~{Schuh}, H.~{Drechsel}, \& U.~{Heber}, 102--109

\bibitem[{{Reffert} {et~al.}(2015){Reffert}, {Bergmann}, {Quirrenbach},
  {Trifonov}, \& {K{\"u}nstler}}]{Reffert2015}
{Reffert}, S., {Bergmann}, C., {Quirrenbach}, A., {Trifonov}, T., \&
  {K{\"u}nstler}, A. 2015, A\&A, 574, A116

\bibitem[{{Reffert} {et~al.}(2006){Reffert}, {Quirrenbach}, {Mitchell},
  {Albrecht}, {Hekker}, {Fischer}, {Marcy}, \& {Butler}}]{Reffert2006}
{Reffert}, S., {Quirrenbach}, A., {Mitchell}, D.~S., {et~al.} 2006, ApJ, 652,
  661

\bibitem[{{Reichert} {et~al.}(in prep.){Reichert}, {Reffert}, {Stock},
  {Trifonov}, {Quirrenbach}, \& {Heeren}}]{Reichert}
{Reichert}, K., {Reffert}, S., {Stock}, S., {et~al.} in prep., A\&A

\bibitem[{{Reimers}(1975)}]{reimers75}
{Reimers}, D. 1975, Memoires of the Societe Royale des Sciences de Liege, 8,
  369

\bibitem[{{Renzini} \& {Buzzoni}(1986)}]{Renzini1986}
{Renzini}, A. \& {Buzzoni}, A. 1986, in Astrophysics and Space Science Library,
  Vol. 122, Spectral Evolution of Galaxies, ed. C.~{Chiosi} \& A.~{Renzini},
  195--231

\bibitem[{{Richichi} {et~al.}(2005){Richichi}, {Percheron}, \&
  {Khristoforova}}]{Richichi2005}
{Richichi}, A., {Percheron}, I., \& {Khristoforova}, M. 2005, A\&A, 431, 773

\bibitem[{{Salpeter}(1955)}]{Salpeter1955}
{Salpeter}, E.~E. 1955, ApJ, 121, 161

\bibitem[{Sato {et~al.}(2007)Sato, Izumiura, Toyota, Kambe, Takeda, Masuda,
  Omiya, Murata, Itoh, Ando, Yoshida, Ikoma, Kokubo, \& Ida}]{Sato2007}
Sato, B., Izumiura, H., Toyota, E., {et~al.} 2007, ApJ, 661, 527

\bibitem[{{Schlaufman} \& {Winn}(2013)}]{SchlaufmanWin2013}
{Schlaufman}, K.~C. \& {Winn}, J.~N. 2013, ApJ, 772, 143

\bibitem[{{Setiawan} {et~al.}(2004){Setiawan}, {Pasquini}, {da Silva},
  {Hatzes}, {von der L{\"u}he}, {Girardi}, {de Medeiros}, \&
  {Guenther}}]{Setiawan2004}
{Setiawan}, J., {Pasquini}, L., {da Silva}, L., {et~al.} 2004, A\&A, 421, 241

\bibitem[{{Shull} \& {Thronson}(1993)}]{Shull1993}
{Shull}, J.~M. \& {Thronson}, H.~A., eds. 1993, {The evolution of galaxies and
  their environment}

\bibitem[{{Skrutskie} {et~al.}(2006){Skrutskie}, {Cutri}, {Stiening},
  {Weinberg}, {Schneider}, {Carpenter}, {Beichman}, {Capps}, {Chester},
  {Elias}, {Huchra}, {Liebert}, {Lonsdale}, {Monet}, {Price}, {Seitzer},
  {Jarrett}, {Kirkpatrick}, {Gizis}, {Howard}, {Evans}, {Fowler}, {Fullmer},
  {Hurt}, {Light}, {Kopan}, {Marsh}, {McCallon}, {Tam}, {Van Dyk}, \&
  {Wheelock}}]{2Mass2006}
{Skrutskie}, M.~F., {Cutri}, R.~M., {Stiening}, R., {et~al.} 2006, ApJ, 131,
  1163

\bibitem[{{Sousa} {et~al.}(2015){Sousa}, {Santos}, {Mortier}, {Tsantaki},
  {Adibekyan}, {Delgado Mena}, {Israelian}, {Rojas-Ayala}, \&
  {Neves}}]{Sousa2015}
{Sousa}, S.~G., {Santos}, N.~C., {Mortier}, A., {et~al.} 2015, A\&A, 576, A94

\bibitem[{{Stello} {et~al.}(2017){Stello}, {Huber}, {Grundahl}, {Lloyd},
  {Ireland}, {Casagrande}, {Fredslund}, {Bedding}, {Palle}, {Antoci},
  {Kjeldsen}, \& {Christensen-Dalsgaard}}]{Stello2017}
{Stello}, D., {Huber}, D., {Grundahl}, F., {et~al.} 2017, MNRAS, 472, 4110

\bibitem[{{Takeda} \& {Tajitsu}(2015)}]{Takeda2015}
{Takeda}, Y. \& {Tajitsu}, A. 2015, MNRAS, 450, 397

\bibitem[{{Tala} {et~al.}(in prep.){Tala}, {Reffert}, \&
  {Quirrenbach}}]{Tala2018}
{Tala}, M., {Reffert}, S., \& {Quirrenbach}, A. in prep., A\&A

\bibitem[{{Tang} {et~al.}(2014){Tang}, {Bressan}, {Rosenfield}, {Slemer},
  {Marigo}, {Girardi}, \& {Bianchi}}]{Tang2014}
{Tang}, J., {Bressan}, A., {Rosenfield}, P., {et~al.} 2014, MNRAS, 445, 4287

\bibitem[{{Torres}(2010)}]{torres2010}
{Torres}, G. 2010, ApJ, 140, 1158

\bibitem[{{Trifonov} {et~al.}(2014){Trifonov}, {Reffert}, {Tan}, {Lee}, \&
  {Quirrenbach}}]{Trifonov2014}
{Trifonov}, T., {Reffert}, S., {Tan}, X., {Lee}, M.~H., \& {Quirrenbach}, A.
  2014, A\&A, 568, A64

\bibitem[{{Turon} {et~al.}(1993){Turon}, {Creze}, {Egret}, {Gomez}, {Grenon},
  {JahreiB}, {Requieme}, {Argue}, {Bec-Borsenberger}, {Dommanget},
  {Mennessier}, {Arenou}, {Chareton}, {Crifo}, {Mermilliod}, {Morin},
  {Nicolet}, {Nys}, {Prevot}, {Rousseau}, {Perryman}, {Arlot}, {Baglin},
  {Barthes}, {Baylac}, {Brosche}, {Burnet}, {Delhaye}, {Dettbarn}, {Erbach},
  {Figueras}, {Fricke}, {Helmer}, {Hemenway}, {Jordi}, {Lampens}, {Lederle},
  {Lub}, {Manfroid}, {Mattci}, {Mazurier}, {Mermilliod}, {Morrison}, {Murray},
  {Oblak}, {Perie}, {Pernier}, {Le Poole}, {Quijano}, {Rapaport}, {Sellier},
  {Torra}, {Tucholke}, {de Vegt}, {Argyle}, {Bacchus}, {Baron}, {Calaf},
  {Cordoni}, {Fabricius}, {Feaugas}, {Fehlberg}, {Florkowski}, {de Geus},
  {Gibbs}, {Hartman}, {Jauncey}, {Johnston}, {Marouard}, {Mekkas}, {Muinos},
  {Nunez}, {Ochsenbein}, {de Orus}, {Paredes}, {Penston}, {Petersen}, {Peyrin},
  {Robin}, {Roman}, {Rossello}, {Schwan}, {Sinachopoulos}, {White},
  {Zacharias}, {Hog}, {Kovalevsky}, {van Leeuwen}, {Lindegren}, {Schutz}, \&
  {Schrijver}}]{Turon1993}
{Turon}, C., {Creze}, M., {Egret}, D., {et~al.} 1993, Bulletin d'Information du
  Centre de Donnees Stellaires, 43

\bibitem[{{van Leeuwen}(2007)}]{vanLeeuwen2007}
{van Leeuwen}, F. 2007, A\&A, 474, 653

\bibitem[{Villani \& Larsson(2006)}]{Villani2006}
Villani, M. \& Larsson, R. 2006, Commun. in Stat. - Theory and Methods, 35,
  1123

\bibitem[{{Villaver} {et~al.}(2014){Villaver}, {Livio}, {Mustill}, \&
  {Siess}}]{Villaver2014}
{Villaver}, E., {Livio}, M., {Mustill}, A.~J., \& {Siess}, L. 2014, ApJ, 794, 3

\bibitem[{{Vrard} {et~al.}(2016){Vrard}, {Mosser}, \& {Samadi}}]{Vrard2016}
{Vrard}, M., {Mosser}, B., \& {Samadi}, R. 2016, A\&A, 588, A87

\bibitem[{{Wilson} {et~al.}(2010){Wilson}, {Hearty}, {Skrutskie}, {Majewski},
  {Schiavon}, {Eisenstein}, {Gunn}, {Blank}, {Henderson}, {Smee}, {Barkhouser},
  {Harding}, {Fitzgerald}, {Stolberg}, {Arns}, {Nelson}, {Brunner}, {Burton},
  {Walker}, {Lam}, {Maseman}, {Barr}, {Leger}, {Carey}, {MacDonald}, {Horne},
  {Young}, {Rieke}, {Rieke}, {O'Brien}, {Hope}, {Krakula}, {Crane}, {Zhao},
  {Carr}, {Harrison}, {Stoll}, {Vernieri}, {Holtzman}, {Shetrone},
  {Allende-Prieto}, {Johnson}, {Frinchaboy}, {Zasowski}, {Bizyaev},
  {Gillespie}, \& {Weinberg}}]{Wilson2010}
{Wilson}, J.~C., {Hearty}, F., {Skrutskie}, M.~F., {et~al.} 2010, in Proc.
  SPIE, Vol. 7735, Ground-based and Airborne Instrumentation for Astronomy III,
  77351C

\bibitem[{{Worthey} \& {Lee}(2011)}]{LeeandWorthey}
{Worthey}, G. \& {Lee}, H.-c. 2011, ApJS, 193, 1

\bibitem[{{Wu} {et~al.}(2011){Wu}, {Singh}, {Prugniel}, {Gupta}, \&
  {Koleva}}]{Wu2011}
{Wu}, Y., {Singh}, H.~P., {Prugniel}, P., {Gupta}, R., \& {Koleva}, M. 2011,
  A\&A, 525, A71

\end{thebibliography}

\end{document}